\documentclass[namedreferences]{SolarPhysics}
\usepackage[optionalrh]{spr-sola-addons} 
\usepackage{epsfig}          
\usepackage{graphicx}        
\usepackage{amssymb} 
\usepackage{color}           
\usepackage{url}             

\newcommand{\etal}{{\it et al.}}
\newcommand{\arcsec}{^{''}}
\newcommand{\degree}{\ensuremath{^\circ}}
\usepackage{txfonts}

\begin{document}

\begin{article}

\begin{opening}

\title{Nature of Quiet Sun Oscillations Using Data from the \textit{Hinode}, TRACE, and SOHO Spacecraft}
\author{G.~R.~\surname{Gupta}$^{1,2,3}$\sep
 S.~\surname{Subramanian}$^{4}$\sep
 D.~\surname{Banerjee}$^{1}$\sep
 M.~S.~\surname{Madjarska}$^{4}$\sep
 J.~G.~\surname{Doyle}$^{4}$ }

\runningauthor{G.~R.~Gupta \etal}
\runningtitle{Nature of Quiet--Sun Oscillations}
\institute{$^{1}$Indian Institute of Astrophysics, Koramangala, Bangalore 560 034, India.
                  email: \url{girjesh@gmail.com}\\
$^{2}$Joint Astronomy Programme, Indian Institute of Science, Bangalore 560 012, India.\\
$^{3}$Presently at Max Planck Institute for Solar System Research, 37191, Katlenburg--Lindau, Germany.\\
$^{4}$Armagh Observatory, College Hill, BT61 9DG, Armagh, N. Ireland.\\}

\begin{abstract}
 
We study the nature of quiet--Sun oscillations using multi-wavelength
observations from TRACE, \textit{Hinode}, and SOHO. The aim is to investigate the
existence of propagating waves in the 
solar chromosphere and  the transition region  via analyzing the
statistical distribution of 
power in different locations, \textit{e.g.} in bright magnetic (network), bright non-magnetic and
dark non-magnetic (inter--network) regions, separately. We use Fourier power and phase--difference 
techniques combined with a wavelet analysis. Two-dimensional Fourier power maps were constructed 
in the period bands 2--4 minutes, 4--6 minutes, 6--15 minutes, and beyond 15 minutes. 
We detect the presence of long--period oscillations with periods between 15 and 30 minutes in bright 
magnetic regions. These oscillations were detected from the chromosphere to the 
transition region. The Fourier power maps show that short--period powers are mainly
concentrated in dark regions whereas long--period powers are concentrated in bright magnetic regions.
 This is the first report of long--period 
waves in quiet--Sun network regions. We suggest that the observed propagating oscillations are due 
to magnetoacoustic waves which can be important for the heating of the solar atmosphere. 

\end{abstract}

\keywords{Chromosphere, Quiet; Transition Region;   Oscillations; MHD
waves}
\end{opening}

\section{Introduction}

Although oscillations in the solar photosphere and chromosphere were first detected in the 1960s, they are not as yet well
understood as in the photosphere. When the solar atmosphere is observed at chromospheric heights (Ca~{\sc ii} H and K lines),
a bright web-like cellular pattern called \textquotedblleft the chromospheric network\textquotedblright\ 
pattern is clearly distinguishable \cite{1904ApJ....19...41H}. 
This Ca~{\sc ii} network pattern overlies the photospheric magnetic--field concentrations clustered 
in the supergranular down-flow lanes \cite{1964ApJ...140.1120S,1975ApJ...200..747S}. Regions outside 
the network pattern are called the inter--network. The distinct properties of these network and 
inter--network regions are subject to intense studies \cite{2008AdSpR..42...86H}. 

At the chromospheric level, five and three--minutes oscillations were found to be associated with the network and
 inter--network regions \cite{1984A&A...130..331D}. \inlinecite{1993ApJ...414..345L}  showed that, only the three--minute
 inter--network oscillations are well correlated with the oscillations in the 
underlying photosphere, unlike the long-period 5--20~minutes network oscillations, which were not correlated
 with the photospheric disturbances. \inlinecite{1997ApJ...486L..63C} observed phase--differences between the
 continuum and chromospheric line intensities and interpreted them as a manifestation of upward--propagating
 waves. Similarly, \inlinecite{2000ApJ...531.1150W} also observed upward--propagating waves
 in the upper chromosphere which drove oscillations in the transition region, thus extending the evidence
 for the upward-propagating waves from the
photosphere up to the base of the corona. \inlinecite{2001ApJ...554..424J} suggested that the 
chromospheric oscillations are primarily in response to the $p$--mode oscillations and are often strongly influenced by 
the photospheric magnetic fields before they could reach the transition region. Recently,
 \inlinecite{2007A&A...461L...1V} proposed that a large fraction of the chromospheric acoustic power at frequencies
 below the acoustic cut-off, residing in the proximity of the network field elements, can directly propagate
 from the underlying photosphere. These results show that network magnetic elements can channel low-frequency
 photospheric oscillations into the chromosphere, thus providing 
a way to input mechanical energy into the upper atmosphere \cite{2004Natur.430..536D,2006ApJ...648L.151J}. 

Areas surrounding the quiet--Sun network elements which lack oscillatory power in the 2--3 minutes range are
 termed as \textquotedblleft magnetic shadows\textquotedblright\  \cite{2001ApJ...554..424J,2001ApJ...561..420M}. 
\inlinecite{2007A&A...461L...1V} showed that a large fraction of the quiet chromosphere, surrounding the
 network regions, is occupied by such magnetic shadows which originate from the fibril-like structures
 observed in the core intensity of the Ca~{\sc ii} line. While studying the magnetic network at the
 boundary of an equatorial coronal hole using data from TRACE 1600 \AA\ and 171 \AA\ passbands, 
\inlinecite{2008A&A...488..331T} found a lack of power at high frequencies 
(5.0--8.3 mHz) and significant power enhancements at low (1.3--2.0 mHz) and intermediate
frequencies (2.6--4.0 mHz) above the magnetic network. Recently, \inlinecite{2010A&A...510A..41K}
found three and five minutes power enhancements around the network, forming power halos at photospheric heights.
 These power halos were replaced by magnetic shadows at chromospheric heights, indicating the existence
 of both upward and downward propagating waves. 

In this work, we investigated the wave properties from a statistical point of view using \textit{Hinode}/SOT, SOHO/MDI,
 and TRACE data. Our focus was mainly on the distribution of Fourier power over the quiet Sun, particularly the
 low--frequency domain. We also investigated the influence of small scale network magnetic fields on wave properties.
 For this, we analyzed the power and phase--difference between
oscillations observed simultaneously in the chromosphere and the low transition region in
bright magnetic regions (network), bright non-magnetic regions, and dark non-magnetic (inter--network)
regions of the quiet Sun. 

\section{Observations and Data Reduction}

\begin{table*}[htbp]
\centering
\begin{tabular}{cccc} 
\hline
\textbf{ } & \textbf{SOT} & \textbf{TRACE}  & \textbf{MDI}\\
\hline
\textbf{Time (UT)} & 11:50\,--\,16:59 & 11:50\,--\,16:59  & 12:00\,--\,15:19 \\
\textbf{Bandpass/Wavelength (\AA)} & Ca~{\sc ii} H & 1550 filter  &
Ni~{\sc i}~6768\\
\textbf{FOV} &  $112\arcsec \times 56\arcsec$ & $384\arcsec\times
384\arcsec$ & $600\arcsec \times 600\arcsec$\\
\textbf{cadence (seconds)} &  $30$ & $35$  & $60$\\
\hline
\end{tabular}
\label{tab:tab1}
\caption{Description of the observation.}
\end{table*}

%
Simultaneous observations obtained near disk centre on 9 April 2007 with the \textit{Solar Optical Telescope} 
(SOT; \opencite{2008SoPh..249..167T}) onboard \textit{Hinode}, with the \textit{Transition Region and Coronal Explorer}
(TRACE; \opencite{1999SoPh..187..229H}), and \textit{Michelson Doppler Imager}
(MDI; \opencite{1995SoPh..162..129S}) onboard SOHO were used in this work.

\begin{figure}[htbp]
\centering
\hspace*{-0.6cm}\includegraphics[angle=90, width=11cm]{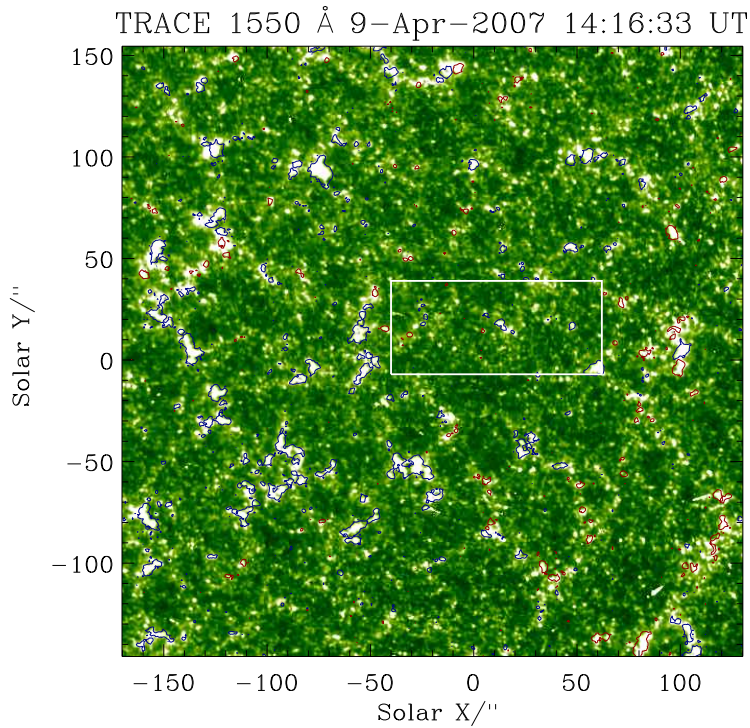}
\caption{The quiet--Sun region observed with the TRACE 1550~\AA\ passband on 9 April 2007. Contour levels
 show the line-of-sight magnetic field strength $\geq|30|$~G obtained from the corresponding MDI
 image. SOT FOV is over-plotted as a rectangular box.}
\label{fig:location}
\end{figure}

TRACE images were obtained in the 1550~\AA\ channel with a field of view (FOV) of $\approx384\arcsec \times 384\arcsec$, 
a pixel size of $0.5\arcsec \times 0.5\arcsec$, and a cadence of 35 seconds. The Ca~{\sc ii} H data from SOT were taken with 
a FOV of $\approx112\arcsec \times 56\arcsec$, a cadence of 30 seconds and a pixel size of 
$0.11\arcsec \times 0.11\arcsec$. Description of the observations is summarized in Table~\ref{tab:tab1}.
The data were reduced using the standard Solar Software package for the 
correction of missing pixels, cosmic--ray hits, CCD bias effects, and flat--field effects. All images from \textit{Hinode} 
and TRACE were converted to SOHO's view (L1). High--resolution photospheric line-of-sight magnetograms were obtained 
with MDI at a cadence of one minute with a pixel size of $0.6\arcsec \times 0.6\arcsec$ and a FOV of 
$\approx620\arcsec \times 300\arcsec$. All of the images were co-aligned with respect to TRACE. We used an available 
procedure for co-aligning TRACE with SOT and MDI data. When images from different instruments are compared, it is 
important to consider the difference in their spatial resolution. First we created an IDL map for all of the 
instruments; which is a structure that contains two-dimensional image data with accompanying pixel coordinate and 
spatial scale information. We used the routine $coreg$\_$map.pro$ available in the SolarSoft library for resizing 
a map with respect to the other, by binning pixels, along with the routine $derot$\_$map.pro$ to de-rotate one map 
with respect to the time of the other map for solar rotation compensation taking into account the roll angle of 
the satellite.  These images were cross correlated and offset coordinates were obtained using the 
$get$\_$corel$\_$offset.pro$ routine. These were then applied to SOT and MDI
images.  After the alignment, Figure~\ref{fig:location} shows the location of the SOT FOV 
on the TRACE 1550~\AA\ image. Contours on the image give the two polarities  of the line-of-sight magnetic 
field obtained by MDI.

\begin{figure}[htbp]
\centering
\hspace*{-0.5cm}\includegraphics[width=12.6cm]{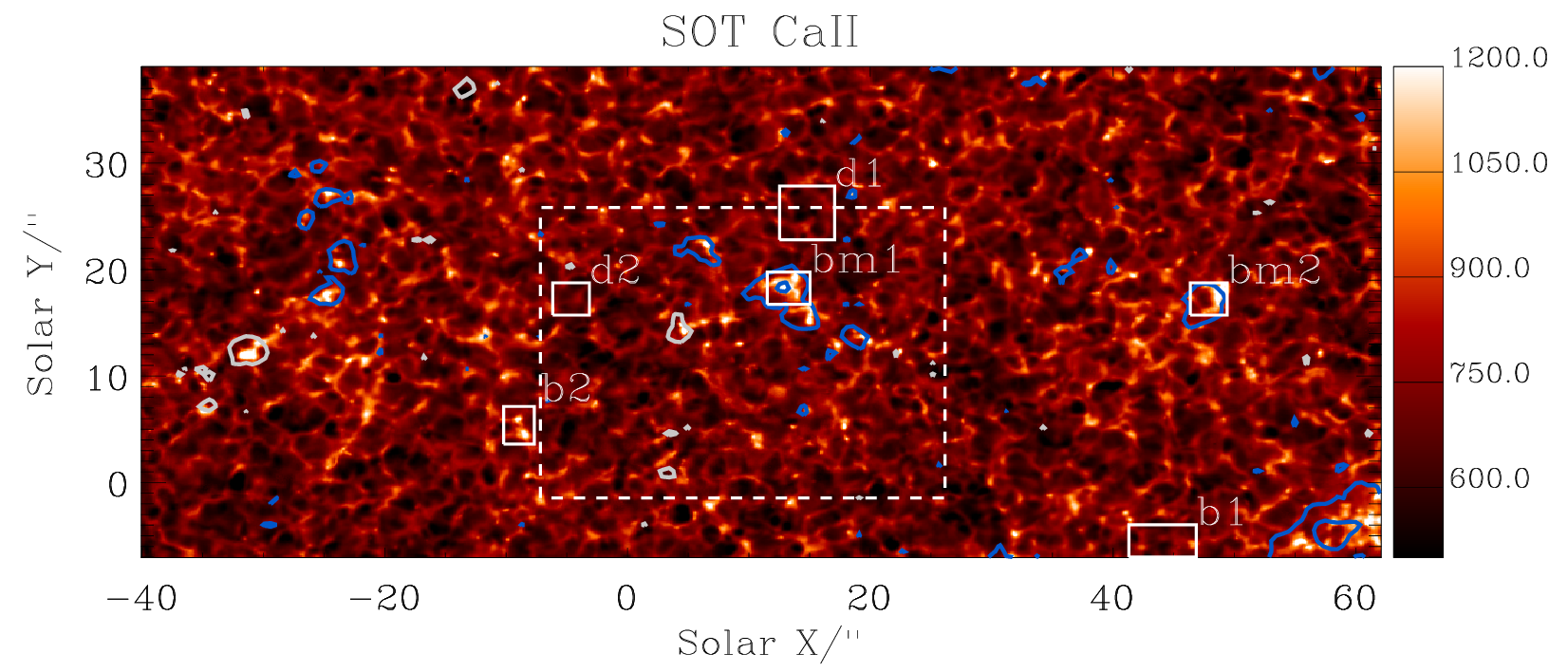}
\hspace*{-0.5cm}\includegraphics[width=12.6cm]{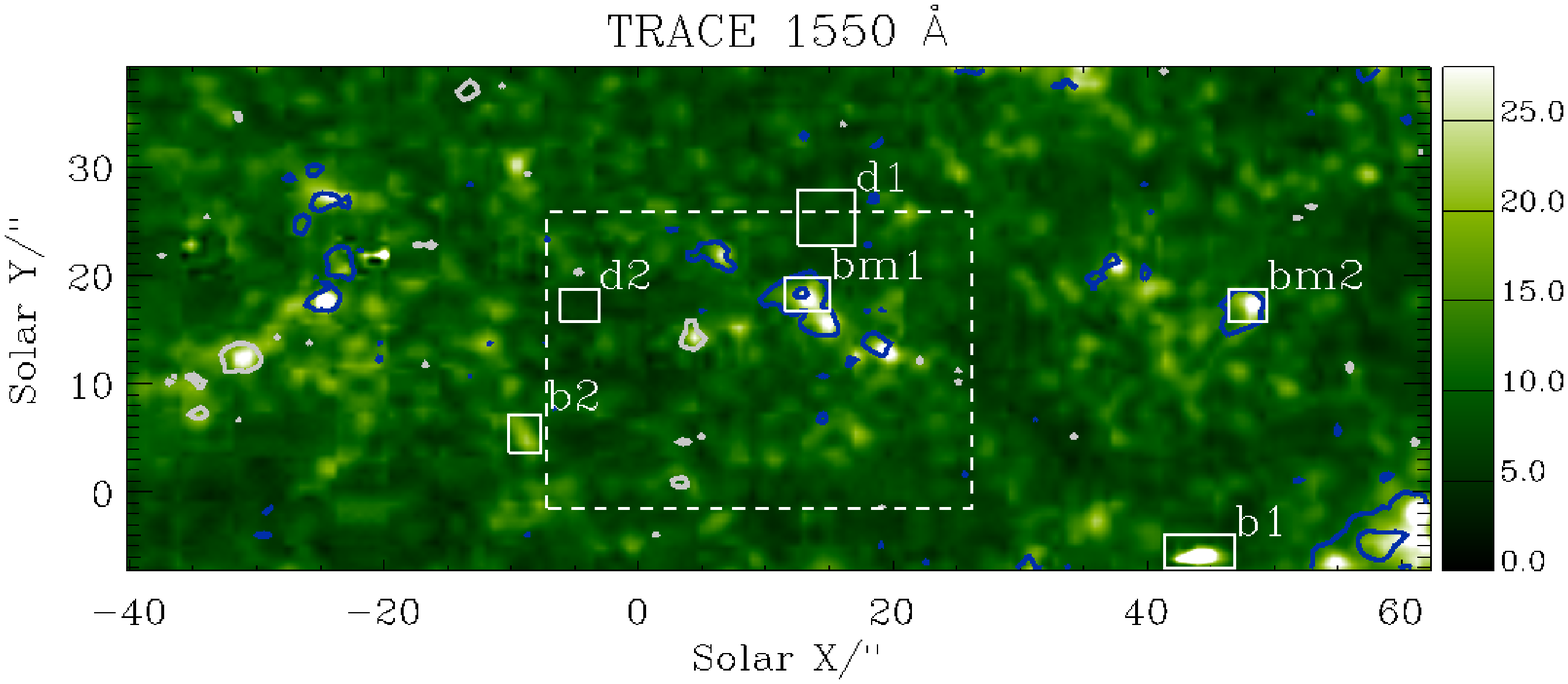}
\caption{Ca~{\sc ii} H (top) and TRACE 1550~\AA\ (bottom) images over-plotted with the corresponding 
 MDI magnetic field contours. Contour levels give the line-of-sight magnetic 
field strength of $\geq|25|$~G. The continuous square boxes (bm1 and bm2), (b1 and b2) and
 (d1 and d2) correspond to bright magnetic, bright non-magnetic and dark non-magnetic
 regions, respectively whereas the dashed box gives the location of the power map as shown in Figure~\ref{fig:power_map}.
 The blue and white contours are the positive and negative polarities of the magnetic field.} 
\label{fig:intensity}
\end{figure}

\section{Data Analysis}

For Fourier power and phase--difference analysis, we choose the longest overlap data between TRACE, and SOT
 (starting at 13:05:10 UT and ending at 15:29:41 UT). We had a few minutes data-gap before (12 minutes 29 seconds)
 and after (30 minutes 50 seconds) the selected SOT observations. During the selected  
time interval, 246 images were obtained in the TRACE 1550~\AA\ passband with a cadence of 35 seconds, 
whereas 286 images were obtained with the Ca~{\sc ii} H broadband filter (SOT) with a cadence of
30 seconds. Data-gaps of a few seconds 
were still present in the data which were linearly interpolated during the analysis. 
We should note that the selected time--series duration for the
present quiet--Sun oscillation study is longer than any previous study and is suitable for
 the detection of oscillations having longer periodicities.

We extracted two sequences of sub-images from the TRACE data set, one having coverage of the MDI FOV (Figure~\ref{fig:location}) 
and another having coverage of the SOT FOV (Figure~\ref{fig:intensity}). Figure~\ref{fig:location} shows a TRACE image 
corresponding to the MDI FOV with the over-plotted MDI magnetic--field contours.
Similarly, Figure~\ref{fig:intensity} shows TRACE and SOT images with the SOT FOV over-plotted with magnetic--field
 contours, clearly illustrating the accuracy of the co-alignment between
the TRACE and SOT data. High resolution movies of TRACE and MDI datasets, one corresponding to the MDI 
FOV and the other corresponding to the SOT FOV, are available as online material 
($mdi_{-}tr_{-}full.mpeg$ and $mdi_{-}tr_{-}small.mpeg$ respectively).  
One can notice that the bright magnetic
elements located in the network intensity--enhanced regions do not move much during the time interval of hours.  

Figures~\ref{fig:location} and \ref{fig:intensity} and the movies (see online) clearly show many bright
magnetic (network) and dark non-magnetic (inter--network) regions. These
distinguishable regions, gave us an opportunity to statistically compare the nature of the Fourier power and 
phase--differences simultaneously  in both regions. We applied a Fourier analysis technique to both TRACE 1550~\AA\ 
and SOT Ca~{\sc ii} H images. A variation of the Fourier power
with frequency were obtained for each spatial pixel. Four different frequency windows, 4.14--8.27 mHz
 (2--4 minutes, to see the 3 minutes period), 2.76--4.02 mHz (4--6 minutes, to
see the 5 minutes period), 1.15--2.64 mHz 
(6--15 minutes, to see the intermediate periods), and up to 1.15 mHz (above 15 minutes,
to see the longer periods), were choosen and the powers were then added together in these frequency
ranges for producing power maps. A subset of these power maps can be found in Figure~\ref{fig:power_map}.

In order to study the influence of the magnetic field on the oscillations, we classified three different
 regions namely, bright magnetic (bm), bright non-magnetic (b), and dark non-magnetic (d) regions
 (Figure~\ref{fig:intensity}). Bright magnetic regions correspond to TRACE 1550~\AA\ 
enhanced intensity locations with a co-spatial magnetic field  $\ge \pm 25$~G as recorded by MDI. Similarly, bright
non-magnetic regions correspond to TRACE 1550~\AA\ 
bright locations  without a co-spatial magnetic field $\ge \pm 25$~G as recorded by MDI (\textit{e.g.} b1 and b2).
 It is also possible that these bright regions were either  associated with a co-spatial magnetic field
 $\le \pm 25$~G or had their footpoints in a region further away from them, \textit{i.e.} not exactly underneath
 the bright locations, as structures on the Sun do not always expand radially. While the rest of the TRACE
 1550~\AA\ dark locations were described as dark non-magnetic regions (\textit{e.g.} d1 and d2).
 Figure~\ref{fig:pwr_region} shows the distribution of power with frequency corresponding to all three regions. 

Now to focus our attention on a few bright magnetic network, bright
non-magnetic locations and dark non-magnetic location, we fix one particular
solar$-X$, namely  $X\approx 14\arcsec$\ and create a distance--time (XT) map from SOT Ca~{\sc ii}
H and the TRACE 1550~\AA\ passbands (Figure~\ref{fig:xt_image}).
A wavelet analysis was performed on the representative bright (network) and dark (inter--network) regions 
as identified in the TRACE 1550~\AA\ XT map (see Figure ~\ref{fig:xt_image}). 
Wavelet results, corresponding to  SOT Ca~{\sc ii} H and TRACE 1550~\AA,  are
shown in Figures~\ref{fig:wav_net} and \ref{fig:wav_int}. A phase
difference analysis was performed between SOT Ca~{\sc ii} H and TRACE
1550~\AA\ passbands. The statistical analysis technique was used to
calculate the time delay between the propagating waves from the Ca~{\sc ii} layers
 to the layers of the TRACE 1550~\AA . 
The  corresponding phase delays are plotted in Figure~\ref{fig:phase_region} for
the different regions. These are discussed in more detail in the next section.

\section{Results and Discussion}

\subsection{Fourier Power Distribution in SOT Ca~{\sc ii} H and  TRACE 1550~\AA\  Passbands}
\label{sec:power}

\begin{figure*}[htbp]
\centering
\hspace*{-1cm}\includegraphics[width=6cm]{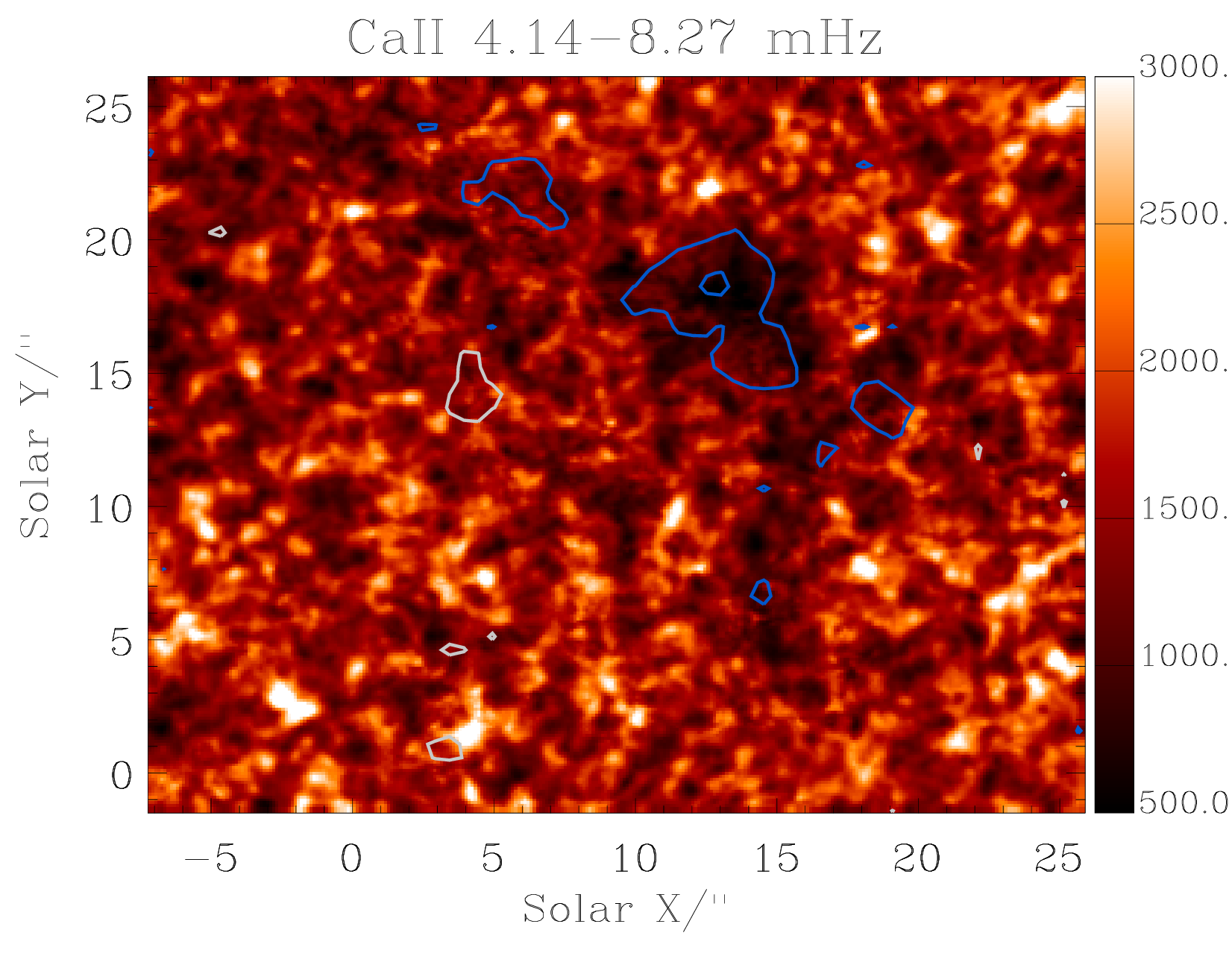}\hspace*{0.5cm}\includegraphics[width=6cm]{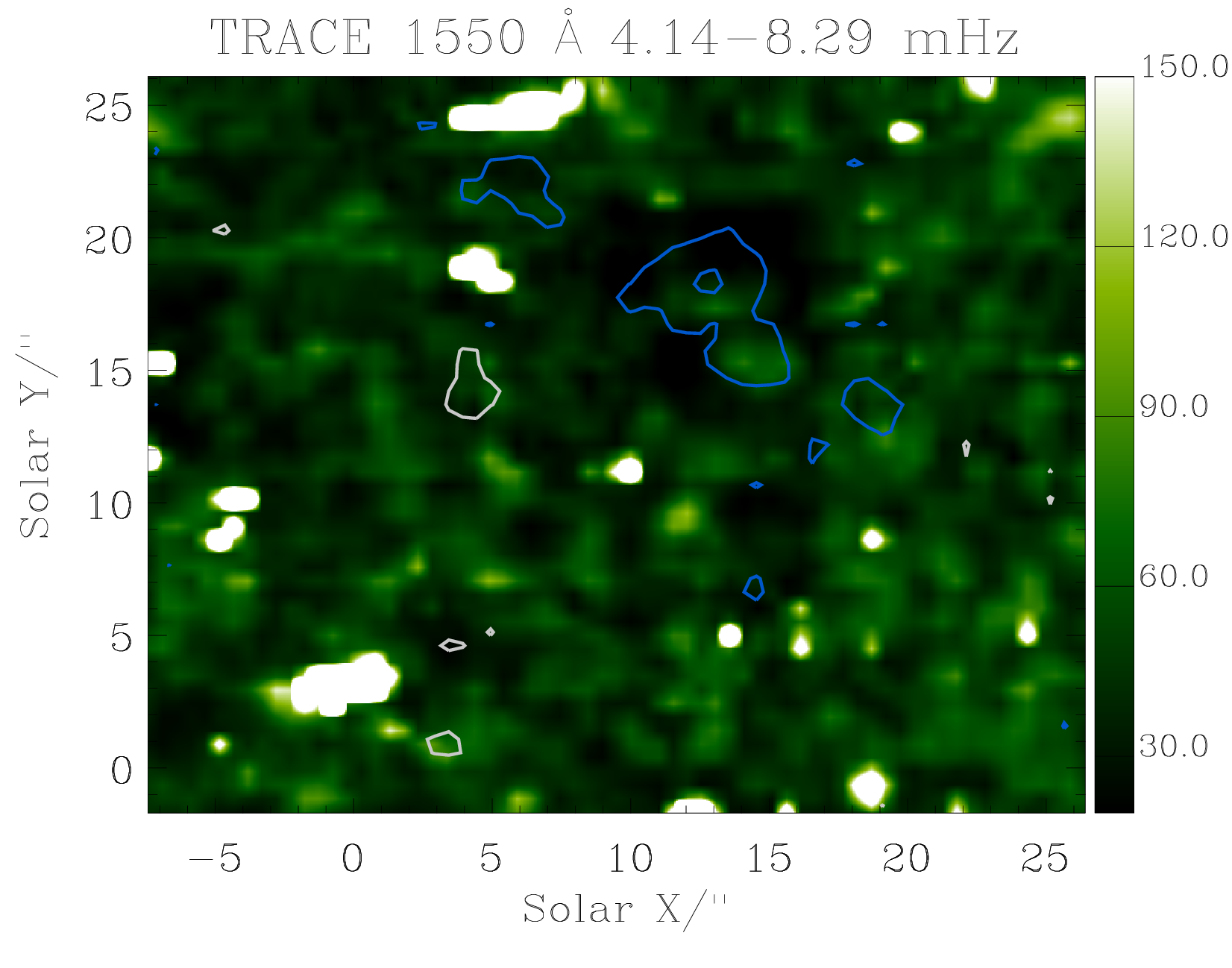}

\hspace*{-1cm}\includegraphics[width=6cm]{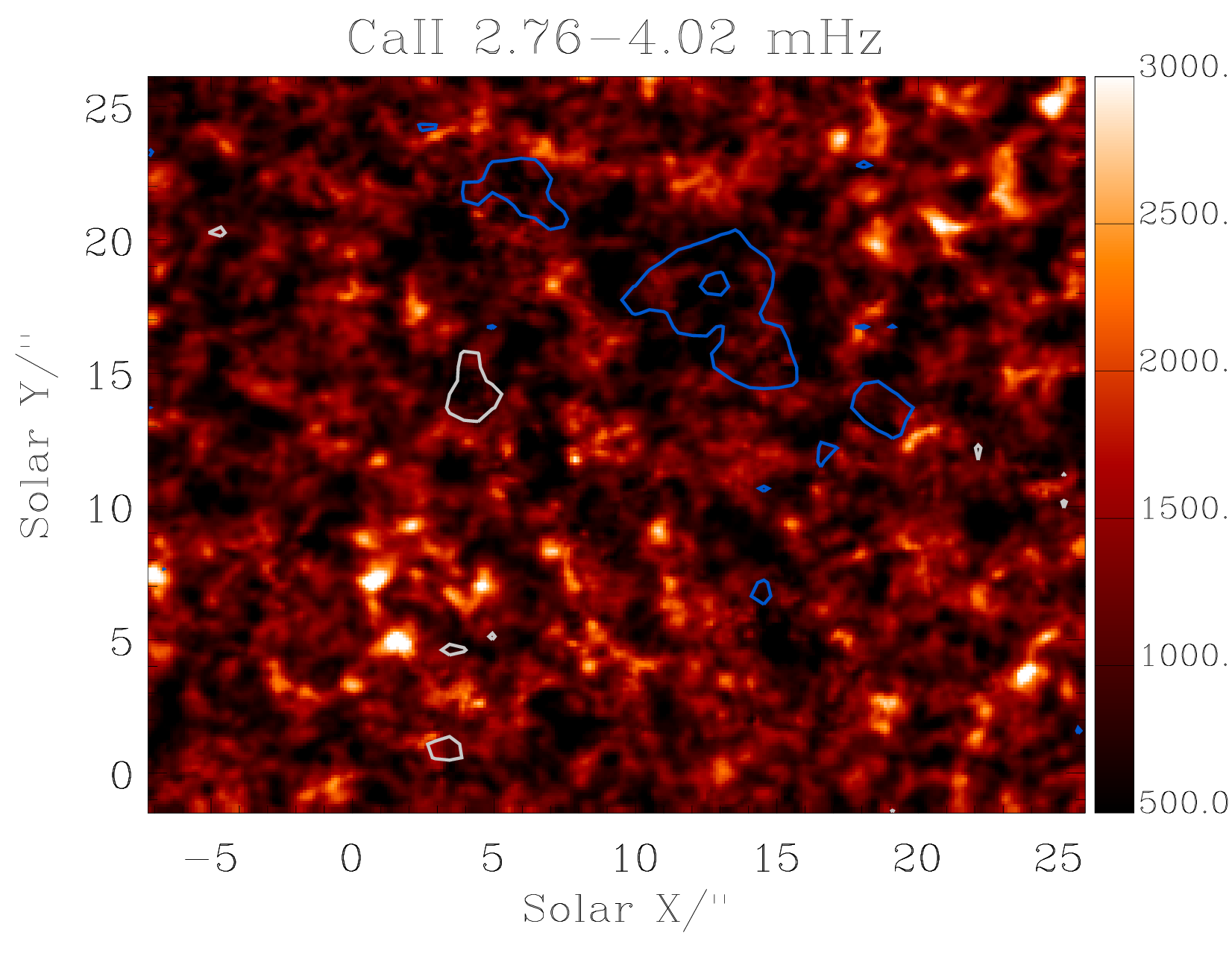}\hspace*{0.5cm}\includegraphics[width=6cm]{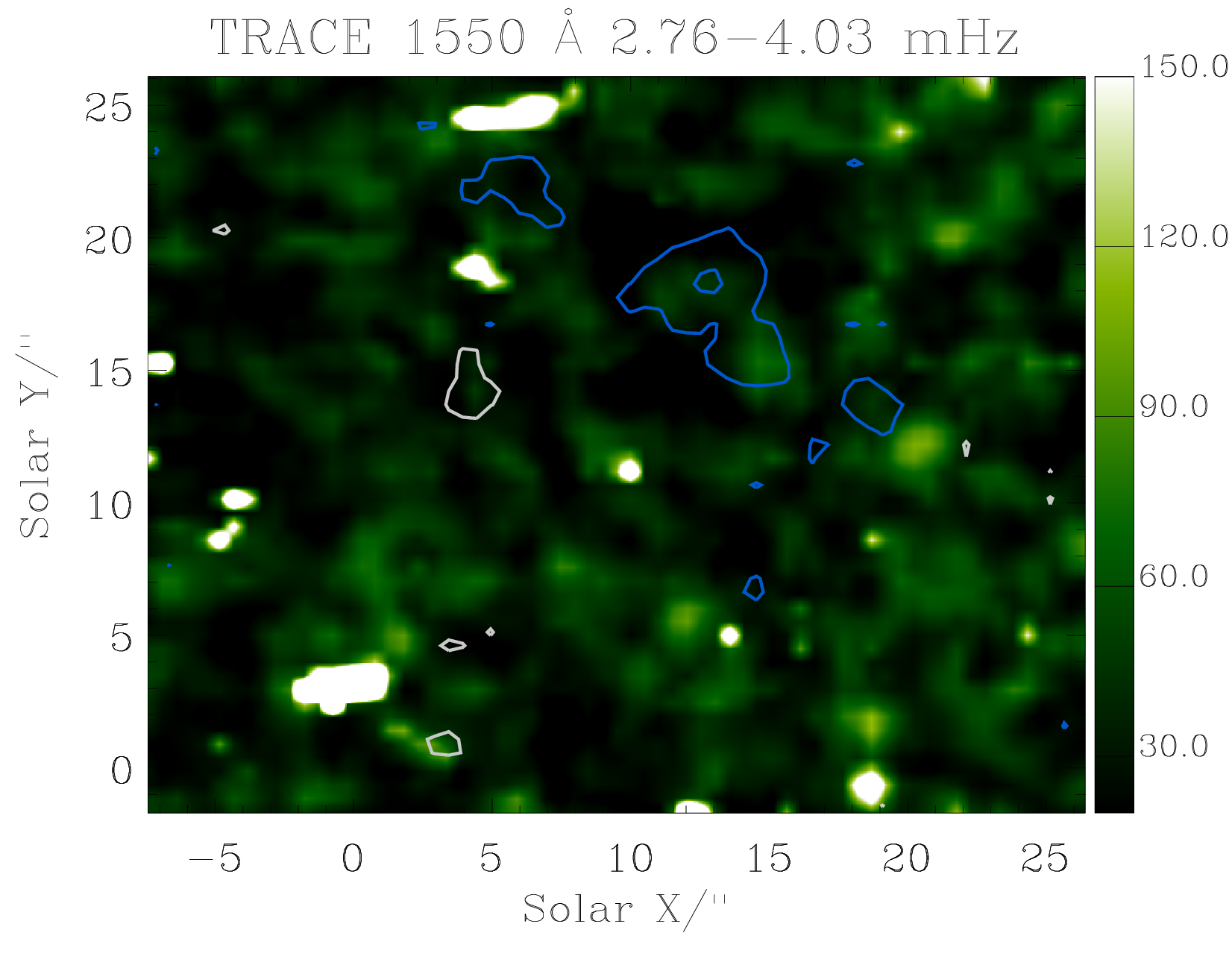}

\hspace*{-1cm}\includegraphics[width=6cm]{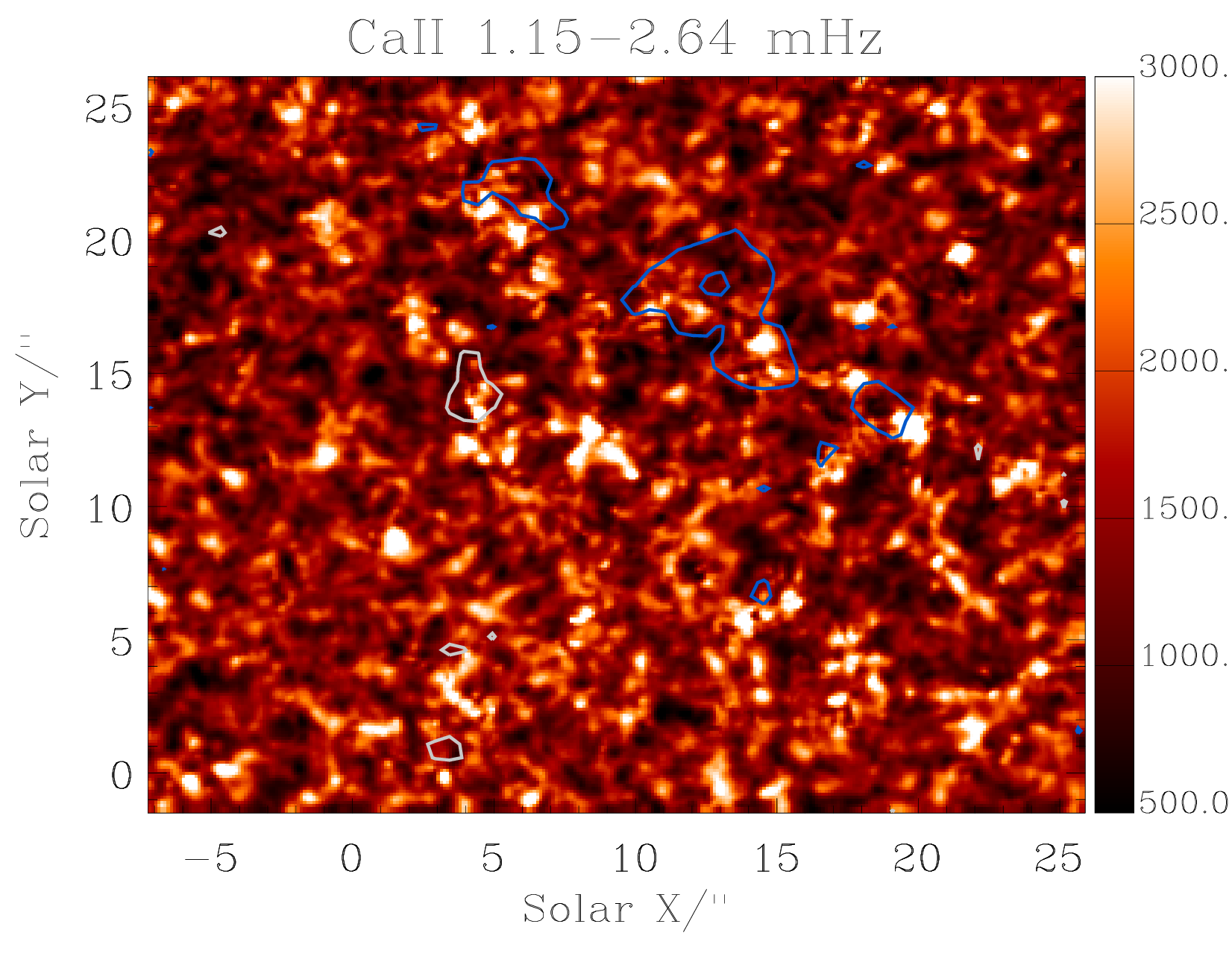}\hspace*{0.5cm}\includegraphics[width=6cm]{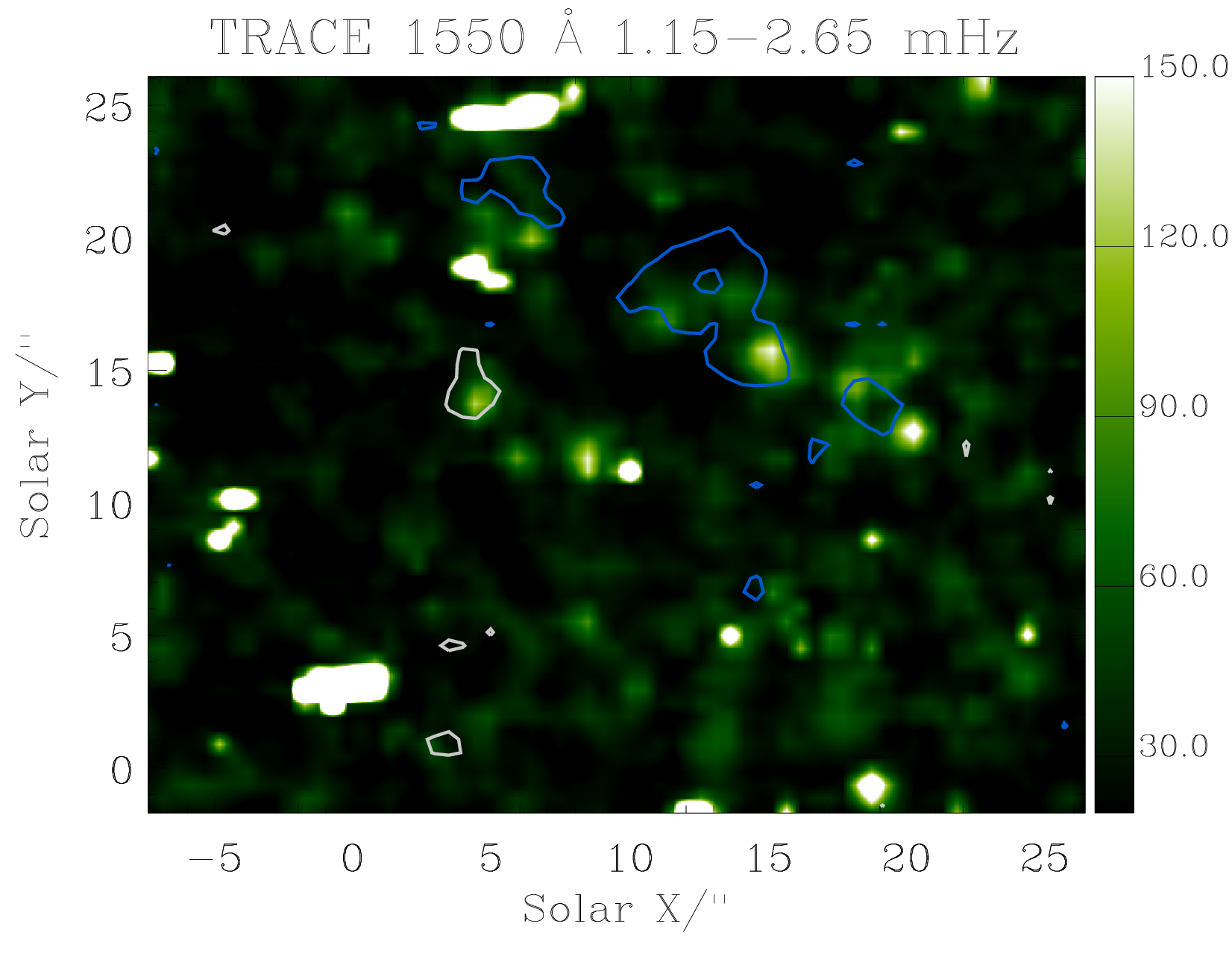}

\hspace*{-1cm}\includegraphics[width=6cm]{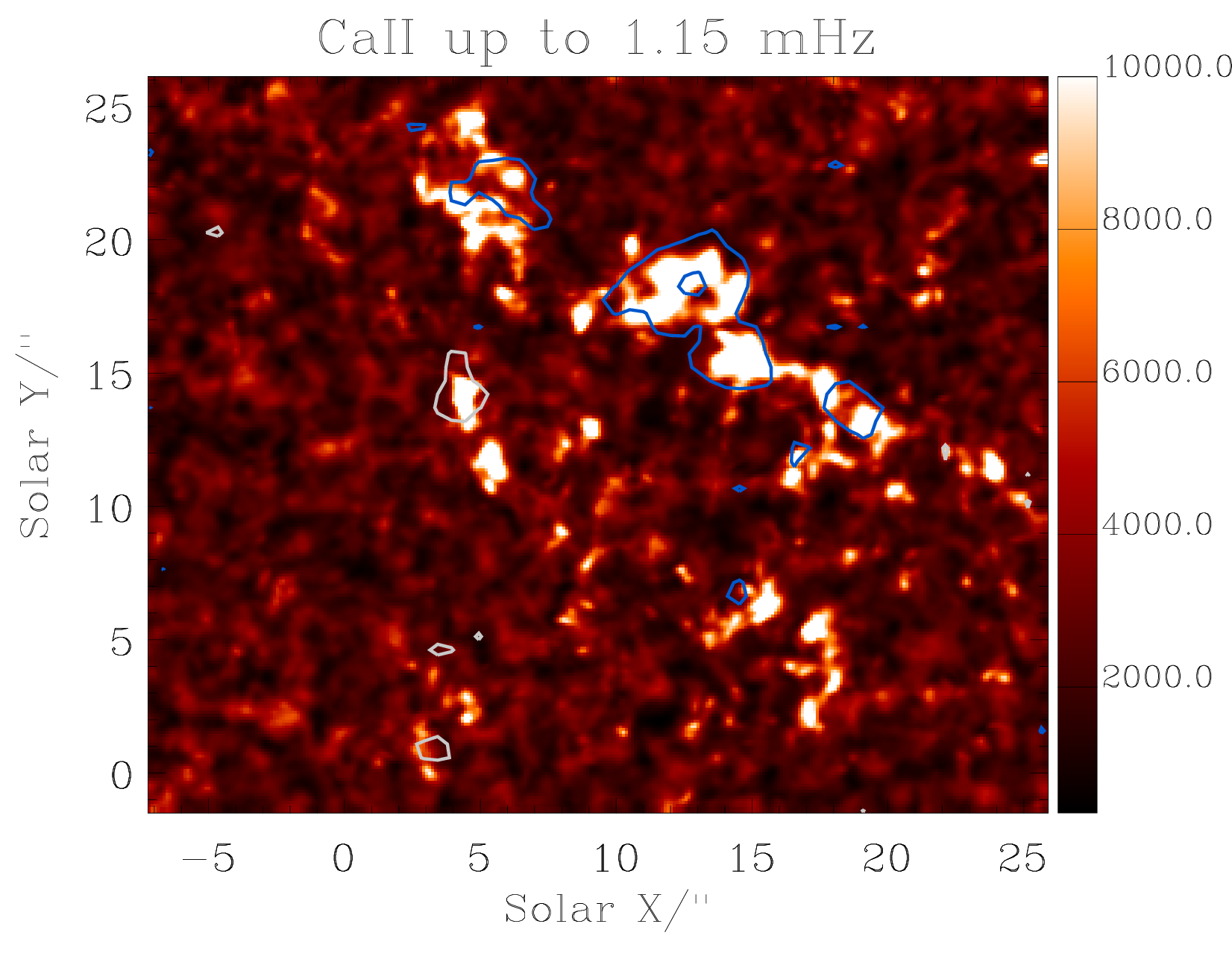}\hspace*{0.5cm}\includegraphics[width=6cm]{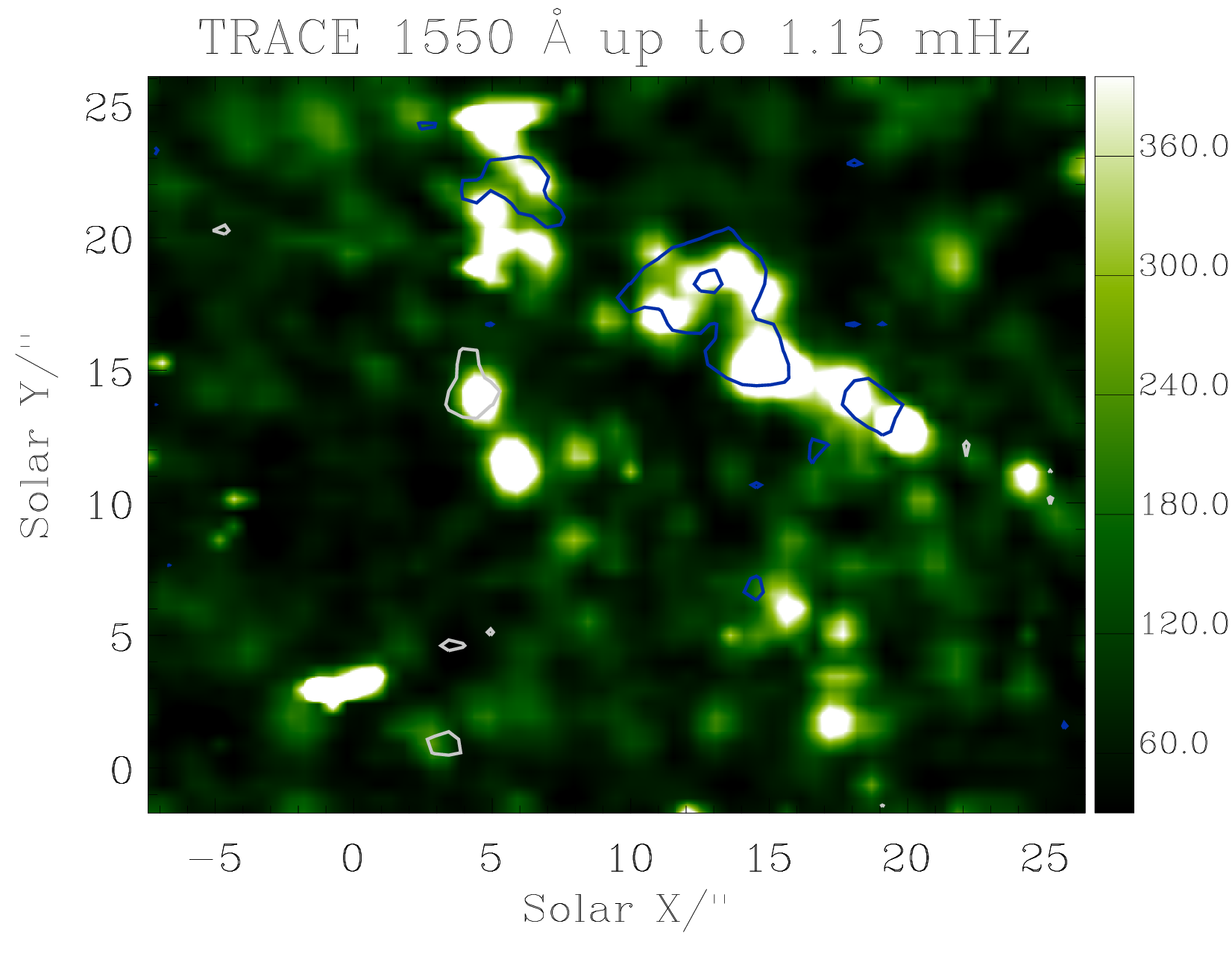}

\caption{Oscillatory power map of Ca~{\sc ii} H (left panels) and TRACE 1550~\AA\ 
(right panels)  on  9 April 2007 in different period ranges as labeled. 
Contour levels give the line-of-sight magnetic--field strength $\geq|25|$~G measured from MDI. 
The blue and white contours are the positive and negative polarities of the magnetic field.} 
\label{fig:power_map}
\end{figure*}

We will now discuss the Fourier power distribution in the SOT Ca~{\sc ii} H and
TRACE 1550~\AA\ passband. In Figure~\ref{fig:intensity}, we show intensity
images in the  SOT Ca~{\sc ii} H and TRACE 1550~\AA\
passbands with the same FOV. To obtain the oscillatory power distribution, a
standard Fourier
power analysis technique was applied to each pixel. While applying this, the standard spatial
resolution of each instrument was retained and the original signal was used without any trend
subtractions or normalization to avoid any kind of bias in the analysis. The resultant 
Fourier power maps in the frequency ranges 4.14--8.27~mHz, 2.76--4.02~mHz, 1.15--2.64~mHz, and 
up to 1.15~mHz were obtained and a subset of these power maps can be found in Figure~\ref{fig:power_map}
for a region marked with dashed boxes in Figure~\ref{fig:intensity}. 

\textit{High frequency range:} the frequency range 4.14--8.27~mHz, which is centered at 3 minutes shows
a lack of power in bright magnetic (network) regions as compared to the neighbouring dark non-magnetic
(inter--network) regions. As  network photospheric flux tubes expand with height 
\cite{1976RSPTA.281..339G}, chromospheric oscillations that appear to be dominant in the inter--network 
can be suppressed above the photospheric network elements as reported by 
\inlinecite{2001ApJ...554..424J}, \inlinecite{2001A&A...379.1052K}, and recently by
\inlinecite{2007A&A...461L...1V} and \inlinecite{2010A&A...510A..41K} for the quiet--Sun regions, and by
\inlinecite{2008A&A...488..331T} at the boundary of an equatorial coronal--hole region. 

\textit{Intermediate frequency range:} the frequency range 2.76--4.02~mHz which is centered at 5~minutes
shows improved power at some bright magnetic (network) regions as seen in the TRACE 1550~\AA\ passband whereas
there is still a lack of power in SOT Ca~{\sc ii} H at many locations. The power in the bright
 regions is almost similar to the power in the high--frequency range. However, reduction in
 the power is observed in dark (inter--network)
 regions in the TRACE 1550~\AA\ passbands as compared to the higher--frequency range. The observed
power in both these ranges can be considered as a mixture of photospheric 5 minutes and chromospheric
3 minutes oscillations \cite{2005Natur.435..919F}.

The frequency range 1.15--2.64~mHz in the TRACE 1550~\AA\ data also shows almost the same behaviour but, here,
SOT Ca~{\sc ii} H shows an enhancement in power in the bright magnetic (network) regions.

\begin{figure*}[htbp]
\centering
\hspace*{-1cm}\includegraphics[width=11cm]{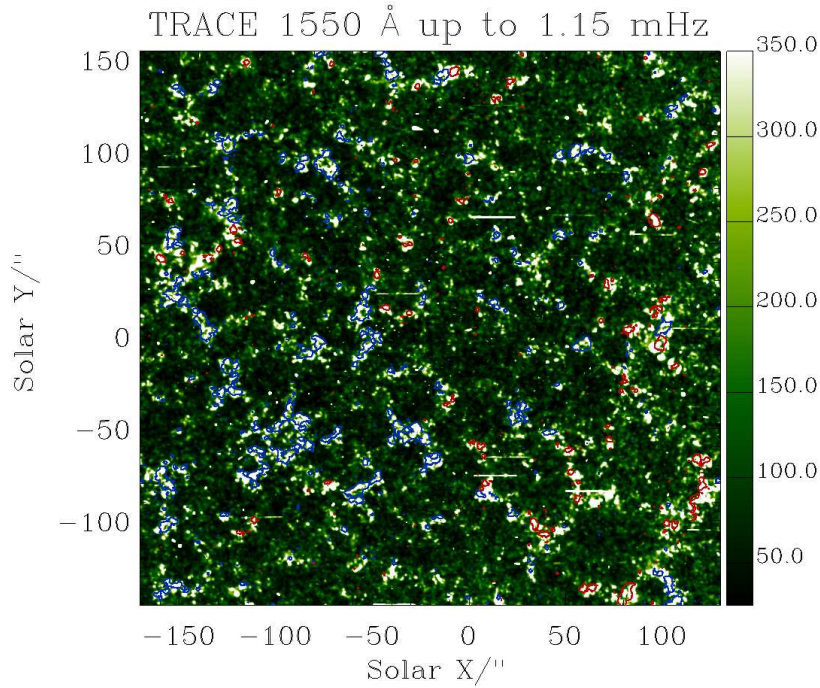}

\caption{Oscillatory power in the period range 15 minutes and above for the full TRACE 1550~\AA\ passband FOV 
 on 9 April 2007. Contour levels give  the line-of-sight magnetic--field strength $\geq|30|$~G measured from MDI. 
The blue and red contours are the positive and negative polarities of the magnetic--field.}
\label{fig:trace_full}
\end{figure*}

\begin{figure*}[htbp]
\centering
\includegraphics[width=8.9cm]{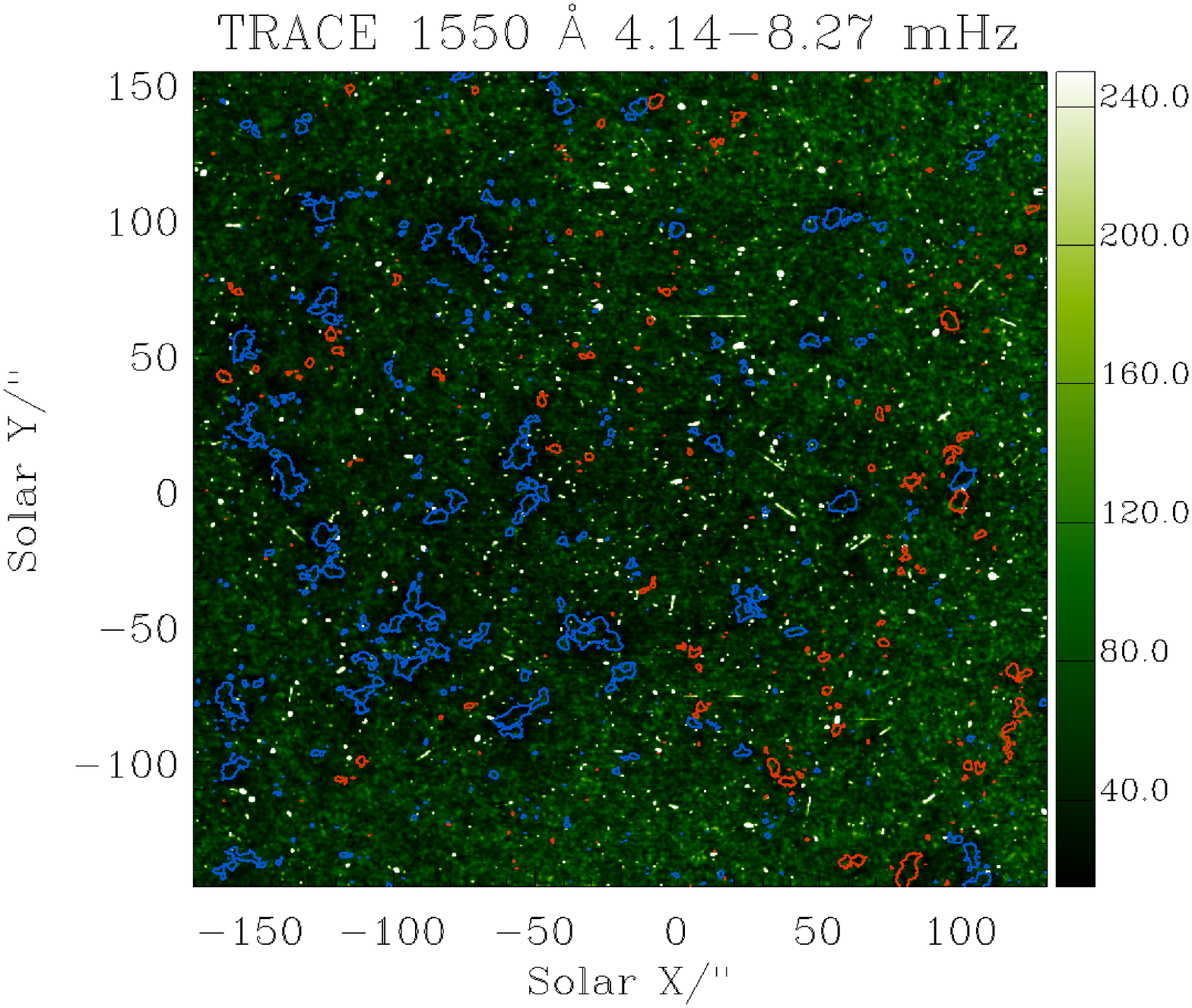}
\vspace*{-0.3cm}\includegraphics[width=8.9cm]{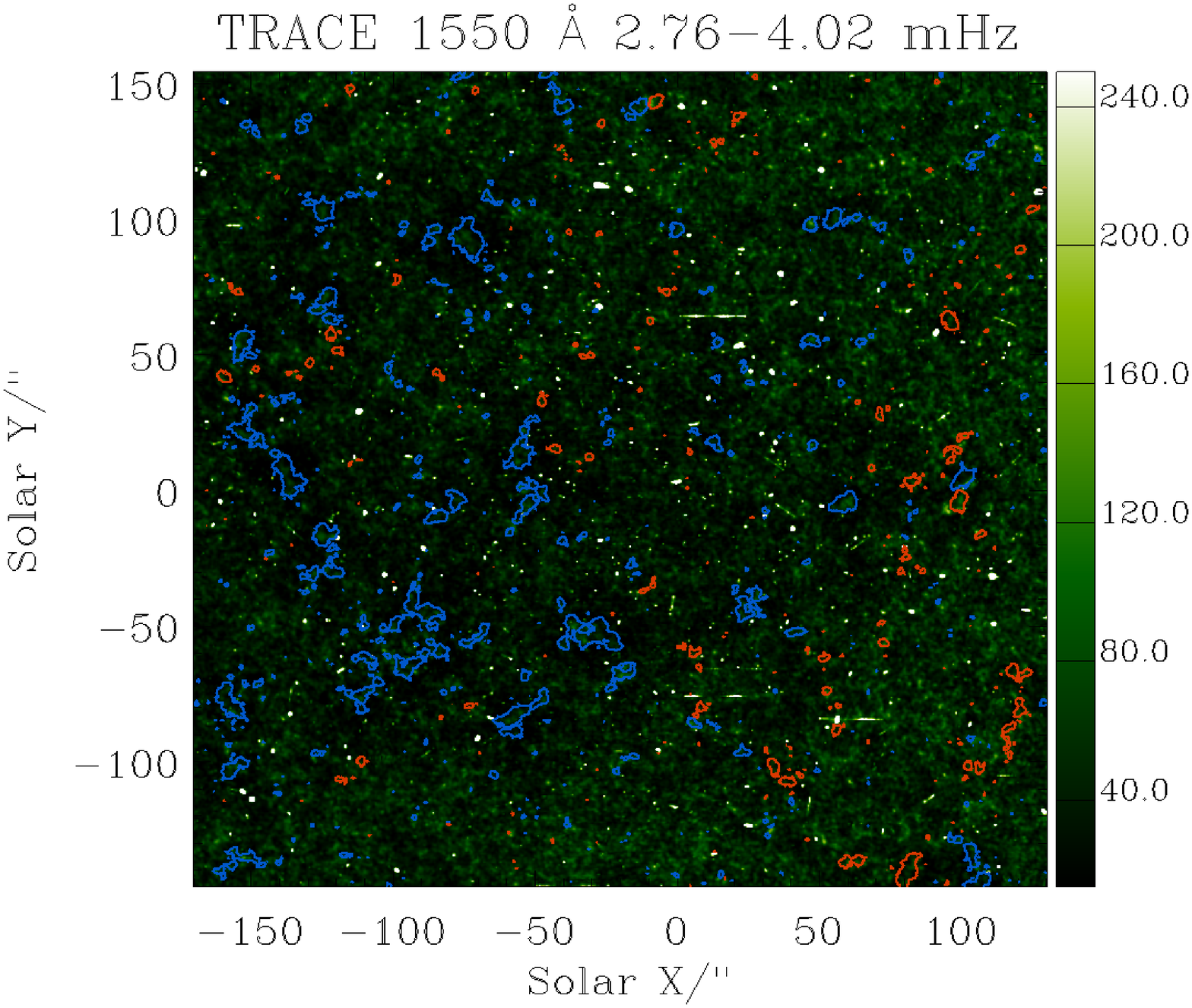}
\vspace*{-0.3cm}\includegraphics[width=8.9cm]{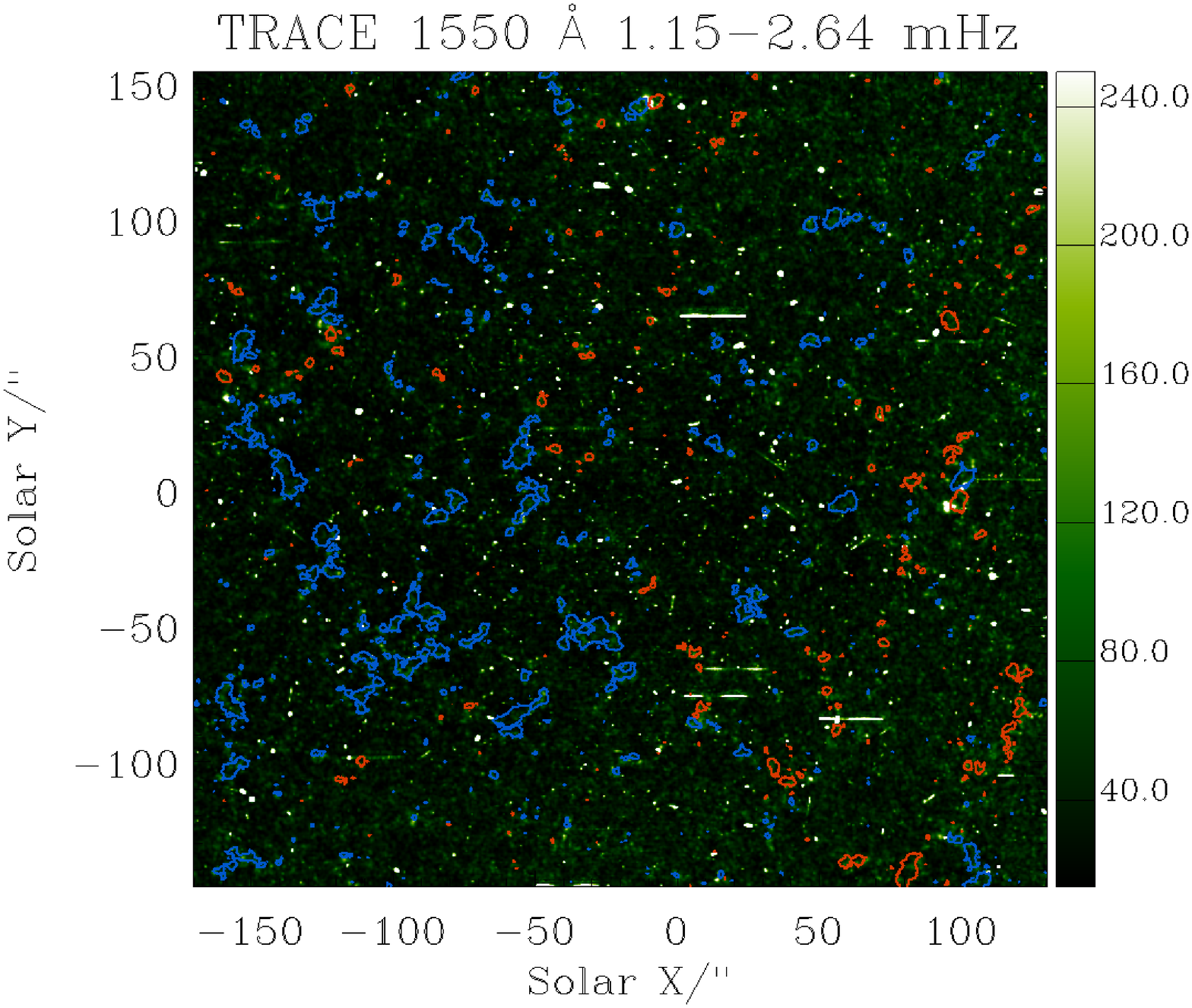}

\caption{Oscillatory power map of TRACE 1550~\AA\ 
 on 9 April 2007 in different frequency ranges as labelled. Contour levels give
 the line-of-sight magnetic--field strength $\geq|30|$~G measured from MDI.
The blue and red contours are the positive and negative polarities of the magnetic--field.} 
\label{fig:power_tr_full}
\end{figure*}

\textit{Low frequency range:} the frequency range up to 1.15~mHz, which covers periods above 15 minutes, 
reveals a map which is completely dominated by a significant power at bright magnetic (network) regions
in both the SOT Ca~{\sc ii} H and the TRACE 1550~\AA\ passbands. This scenario appears much cleaner when observed
in the full TRACE 1550~\AA\ FOV and is shown in Figure~\ref{fig:trace_full}. The long--period powers are
 completely concentrated in network regions which give the impression of power halos and are similar to those
 seen by \inlinecite{2007A&A...461L...1V}, \inlinecite{2008A&A...488..331T}, and 
\inlinecite{2010A&A...510A..41K} for relatively shorter periods and at lower solar atmospheric heights.
The above discussed observations appear much cleaner for high-- and intermediate--frequency ranges when seen in the TRACE
full FOV in Figure~\ref{fig:power_tr_full}. 

The significant power at these low and intermediate frequencies can be explained with the inclined
magnetic--field lines at the boundaries of network structure which provide
\textquotedblleft magnetoacoustic portals\textquotedblright\ through which low--frequency (<5~mHz)
magnetoacoustic waves can propagate into the solar chromosphere \cite{2006ApJ...648L.151J}. 
These waves might provide  a significant source of energy for balancing the radiative losses of the
ambient solar chromosphere.

\begin{figure*}[htbp]
\centering
\includegraphics[width=5.5cm]{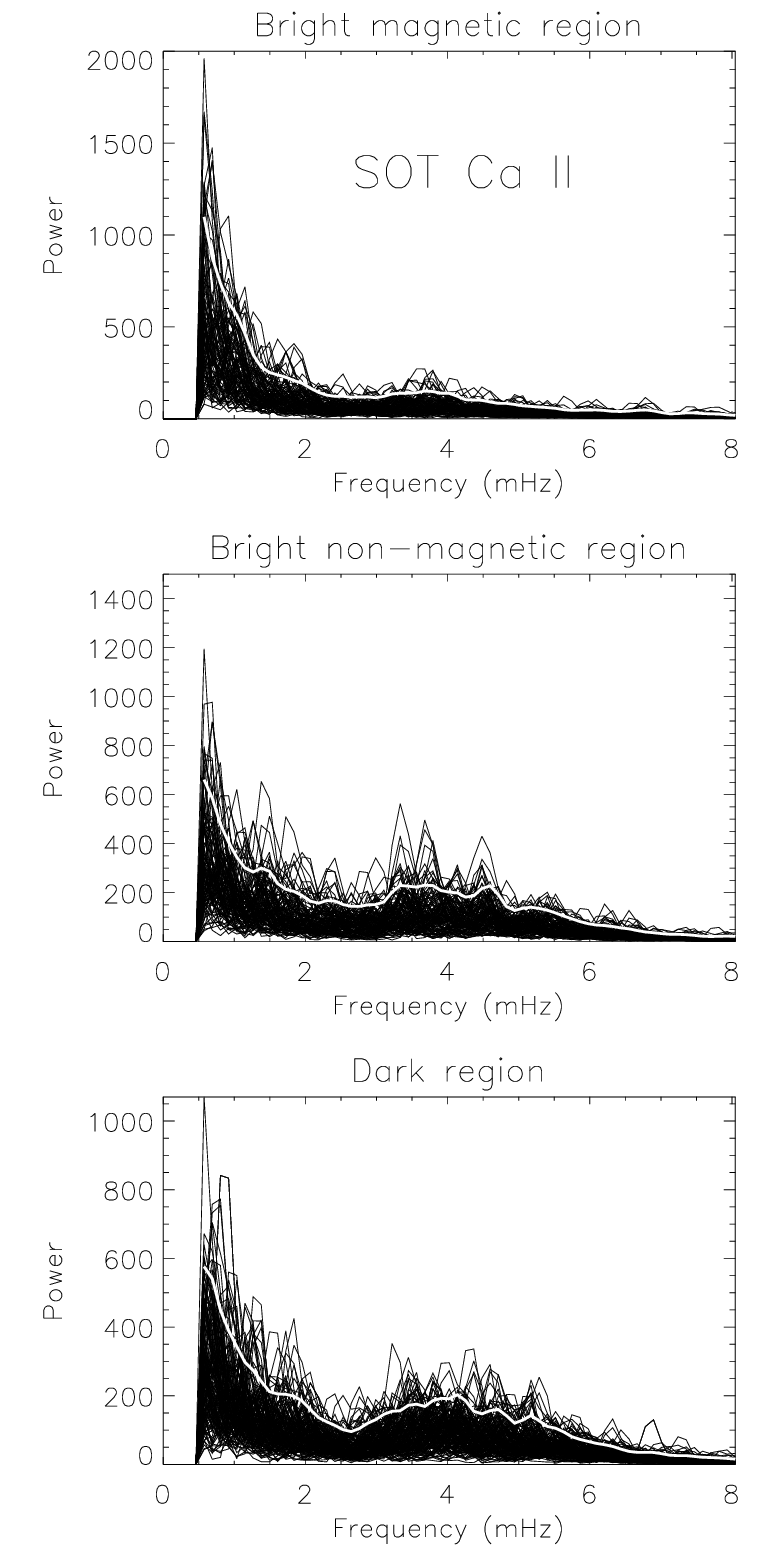}\includegraphics[width=5.5cm]{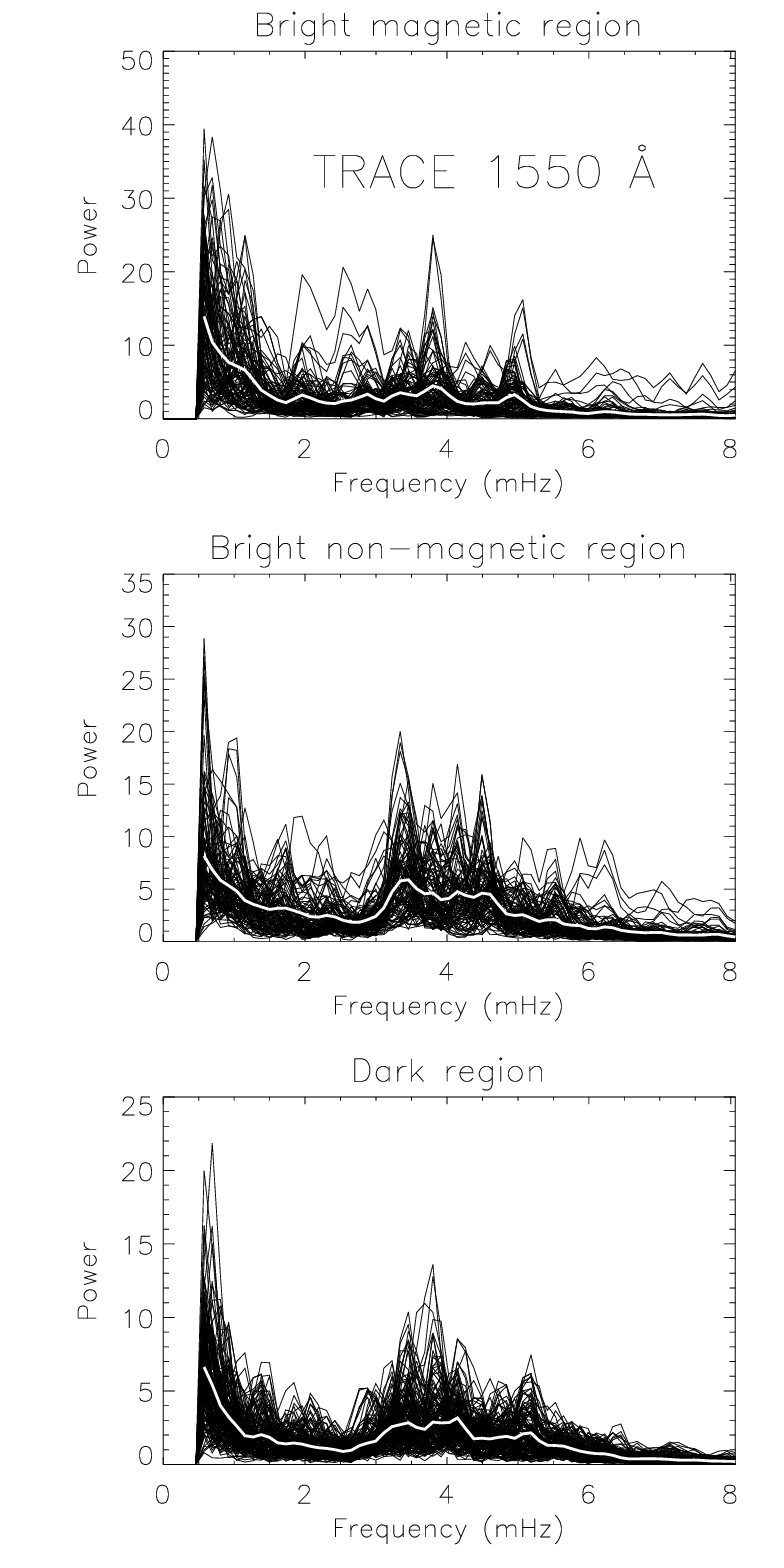}\\
\caption{Fourier power distribution of Ca~{\sc ii} H (left) and TRACE 1550~\AA\ (right) for the few selected
 bright magnetic (bm1 and bm2), bright non-magnetic (b1 and b2) and dark 
non-magnetic (d1 and d2) regions as indicated in Figure~\ref{fig:intensity}.
Overplotted white lines indicate the average Fourier power with respect to frequency.} 
\label{fig:pwr_region}
\end{figure*}

Figure~\ref{fig:pwr_region} shows the typical Fourier power corresponding to each spatial pixel
 within the specific
regions (bm1, bm2, b1, b2, d1, and d2) marked as boxes in Figure~\ref{fig:intensity}, for both the
SOT Ca~{\sc ii} H (left panels) and TRACE 1550~\AA\ (right panels) passbands. 
Here, we filtered out the low--frequency power with typically periods above 40 minutes which may arise
 due to the orbital effect of the spacecraft/instruments. From this figure, it
is clear that low frequency oscillations completely dominate the power distribution in bright
magnetic regions (top panels of Figure~\ref{fig:pwr_region}). Power at other frequencies mainly
between 3--5~mHz begin to appear in comparison to low--frequency power in the bright non-magnetic
 regions (middle panels of Figure~\ref{fig:pwr_region}). 
Whereas power in both frequency ranges become comparable  only in the dark non-magnetic regions. Thus this indicates that the low frequency
power is present in all regions and high--frequency powers between 3 and 5~mHz are only comparable to 
them in dark non-magnetic regions (inter--network). These finding are different from the findings of
\inlinecite{1993ApJ...414..345L} who found low--frequency power only in network regions and high--frequency
 ($\approx5$~mHz) power in inter--network regions.

\begin{figure*}[htbp]
\centering
\hspace*{-0.7cm}\includegraphics[width=11cm]{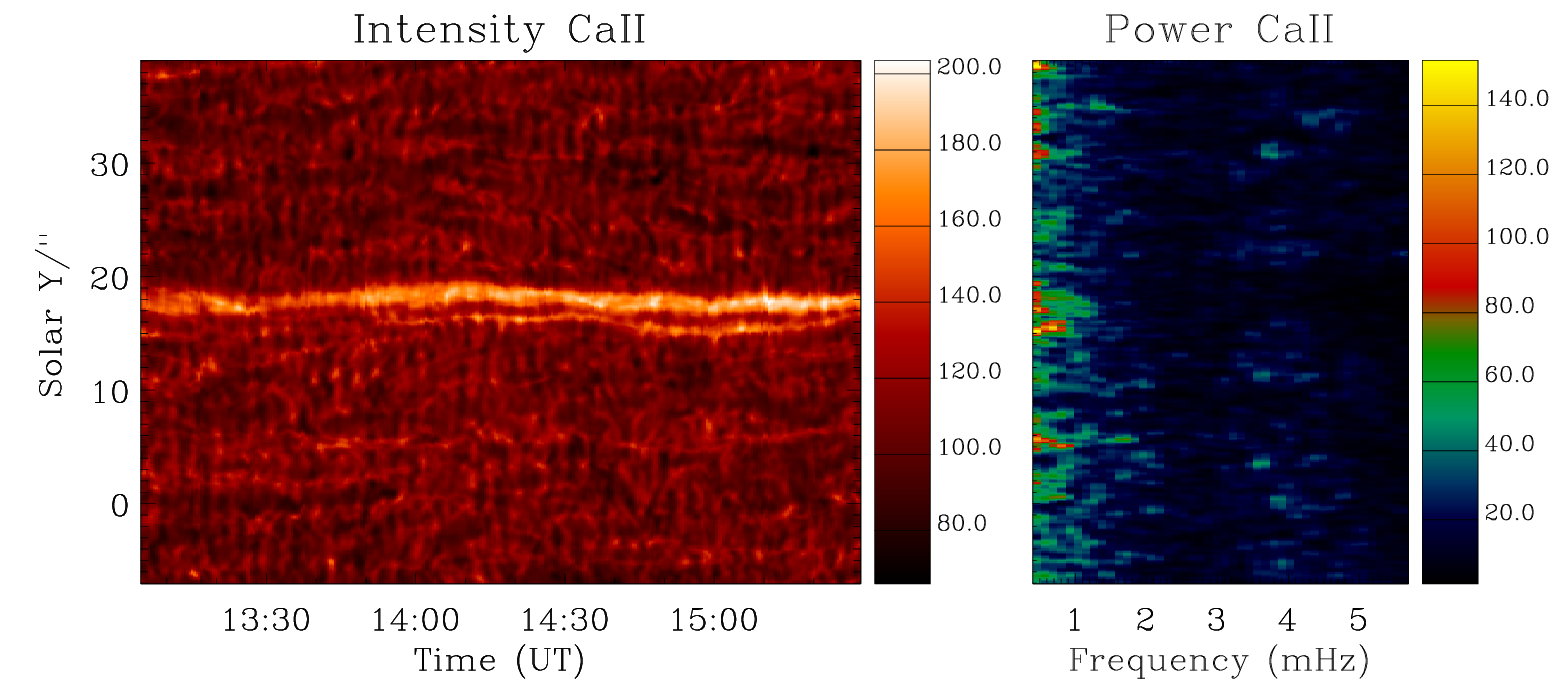} 
\hspace*{-0.7cm}\includegraphics[width=11cm]{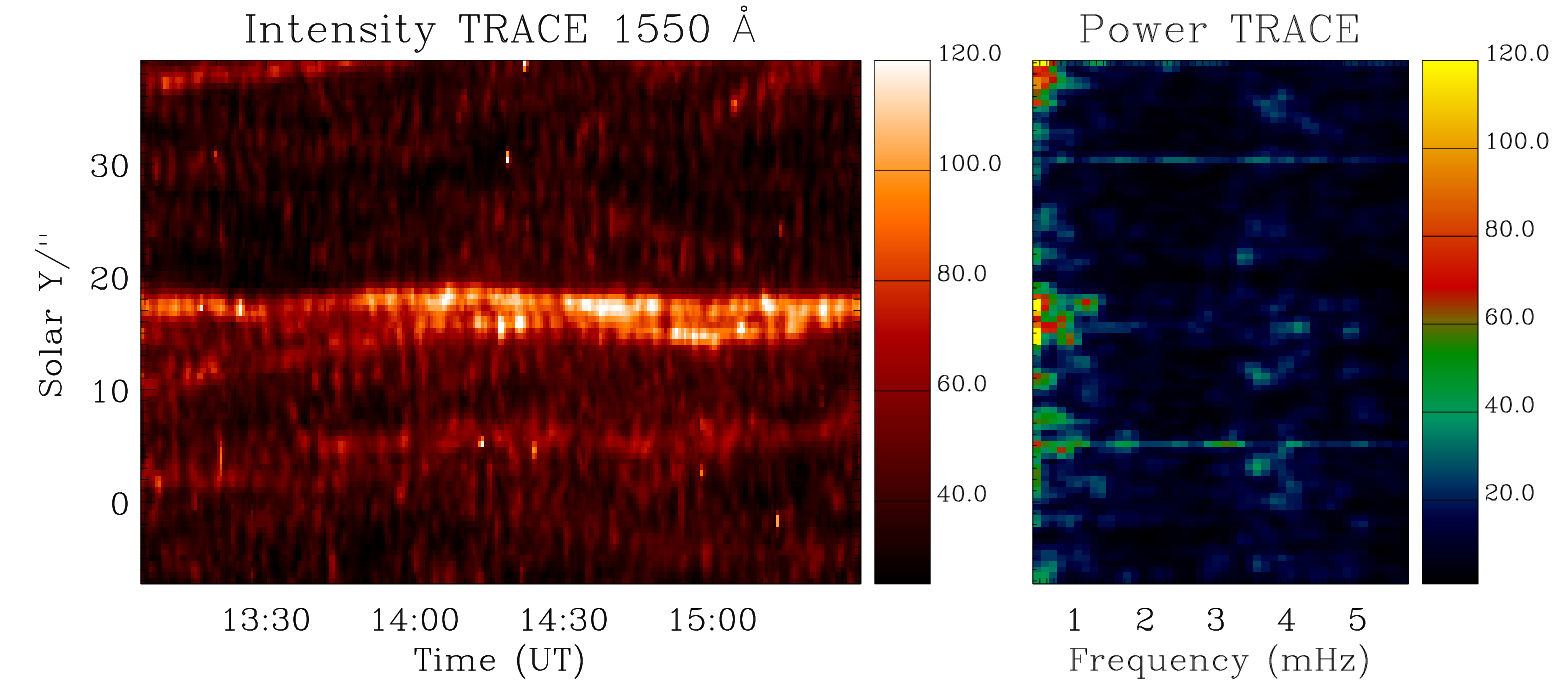} 
\caption{XT intensity map and respective Fourier power map of Ca~{\sc ii} H
 (top) obtained from SOT BFG images and TRACE 1550~\AA\
 (bottom) images  at solar$-X\approx 14\arcsec$.}
 \label{fig:xt_image}
\end{figure*}

\subsection{Wavelet Analysis in Network and Inter--network Regions}

\begin{figure*}[htbp]
\centering
\hspace*{-0.8cm}\includegraphics[angle=90,
width=6.5cm]{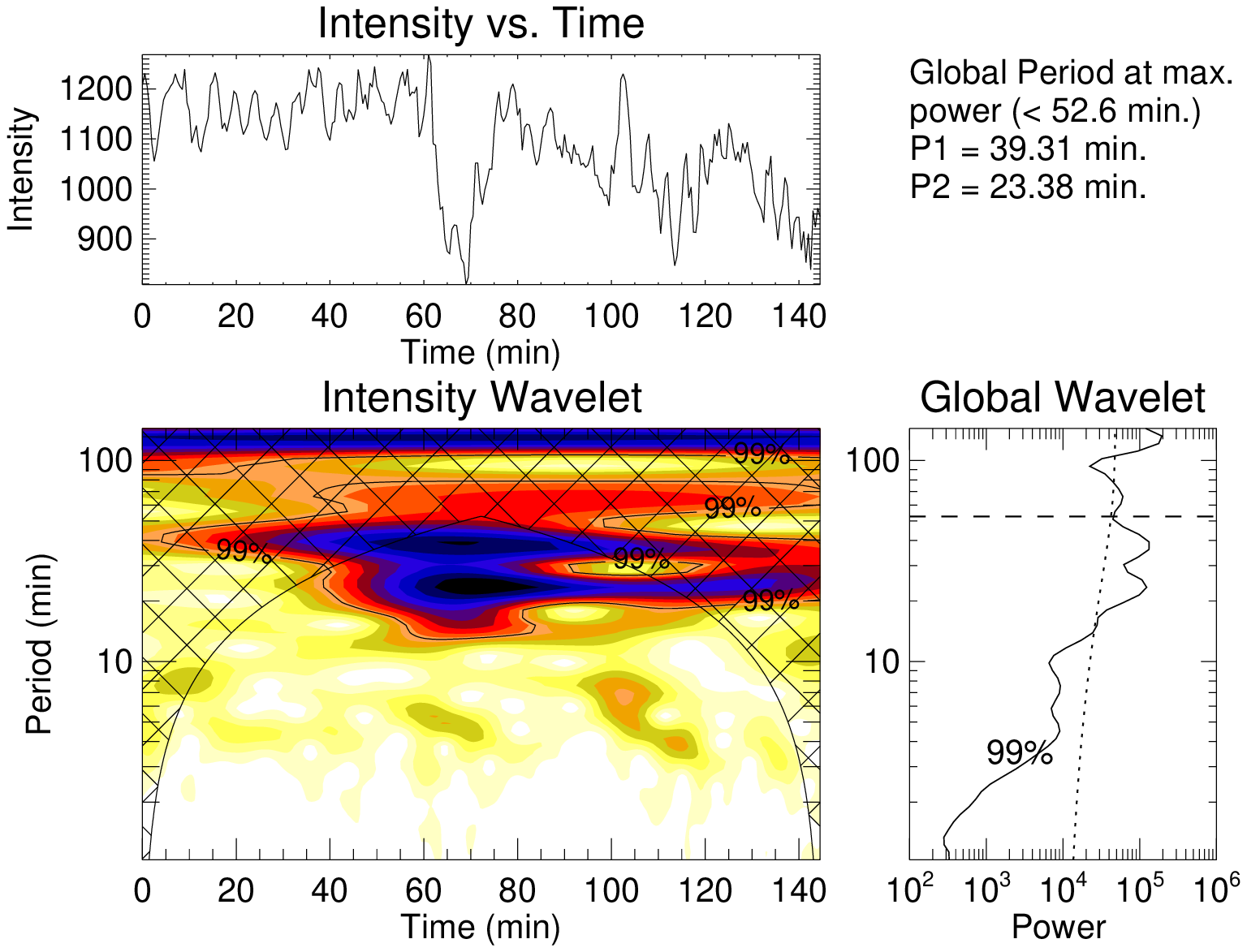}\includegraphics[angle=90,width=6.5cm]{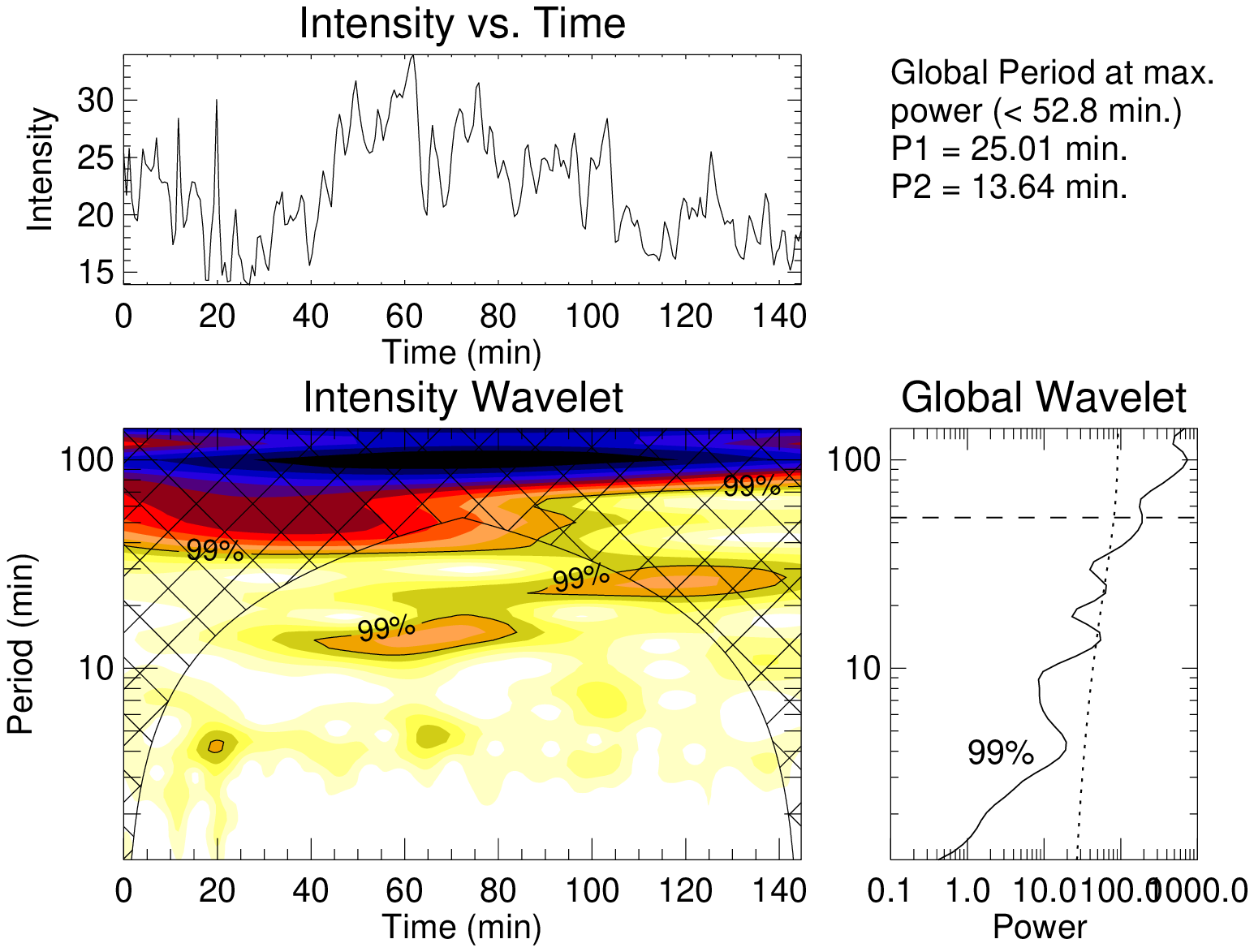}
 \caption{The wavelet results for the network location as identified from TRACE 1550~\AA\ XT map at
 solar$-Y\approx 18\arcsec$ and solar$-X\approx 14\arcsec$ obtained from light
curves of Ca~{\sc ii} H (left) and TRACE 1550~\AA\ (right). In each set, the top panels show the original
light curve. The bottom left panels show the color--inverted wavelet power spectrum
 with $99~\%$ confidence level contours while
 the bottom right panels show the global (averaged over time) wavelet power spectrum with $99\%$
 global confidence level drawn. 
 The periods P1 and P2 at the locations of the first two maxima in the global wavelet spectrum are 
 printed above the global wavelet spectrum.}
 \label{fig:wav_net}
\end{figure*}

\begin{figure*}[htbp]
\centering
\hspace*{-0.8cm}\includegraphics[angle=90,
width=6.5cm]{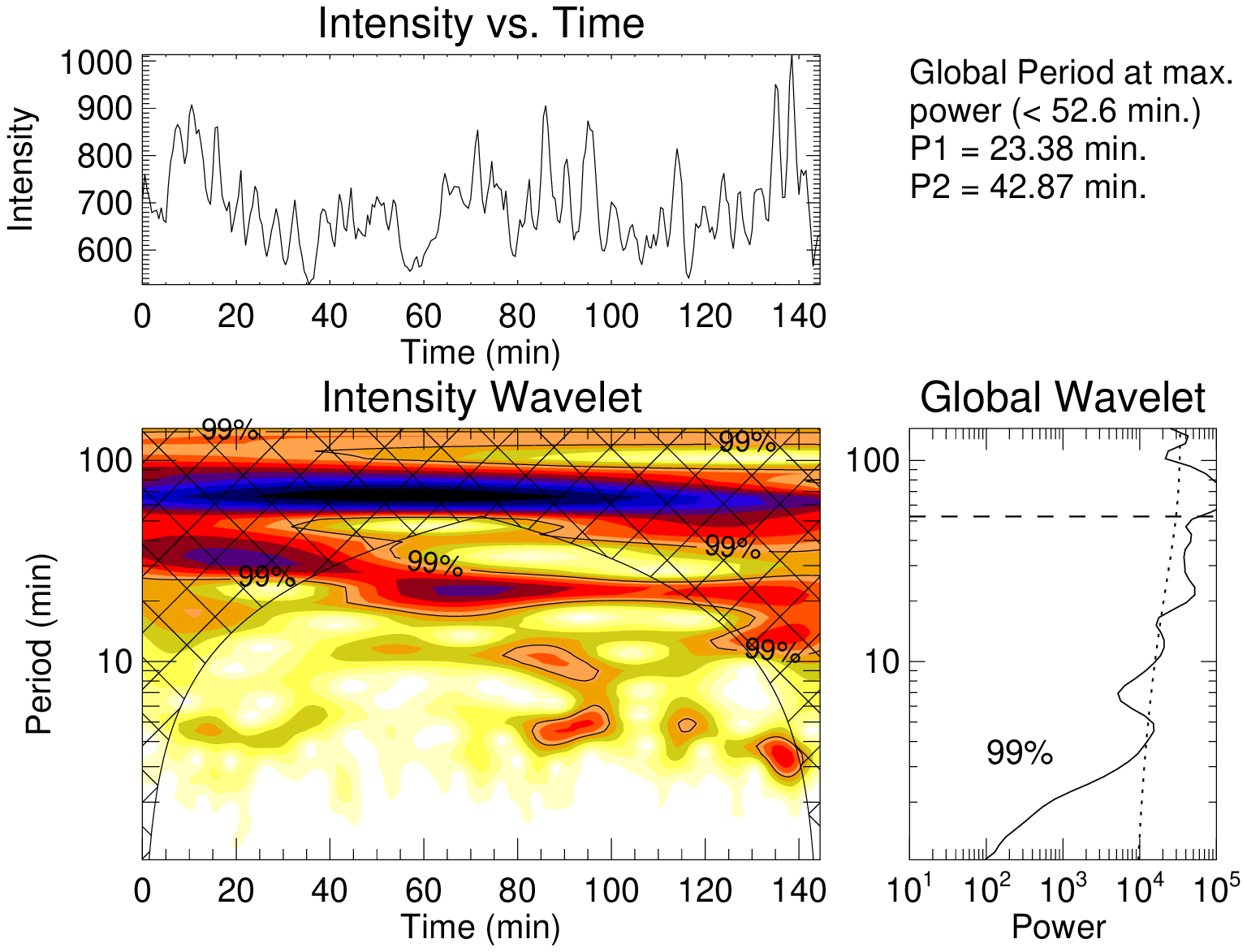}\includegraphics[angle=90,width=6.5cm]{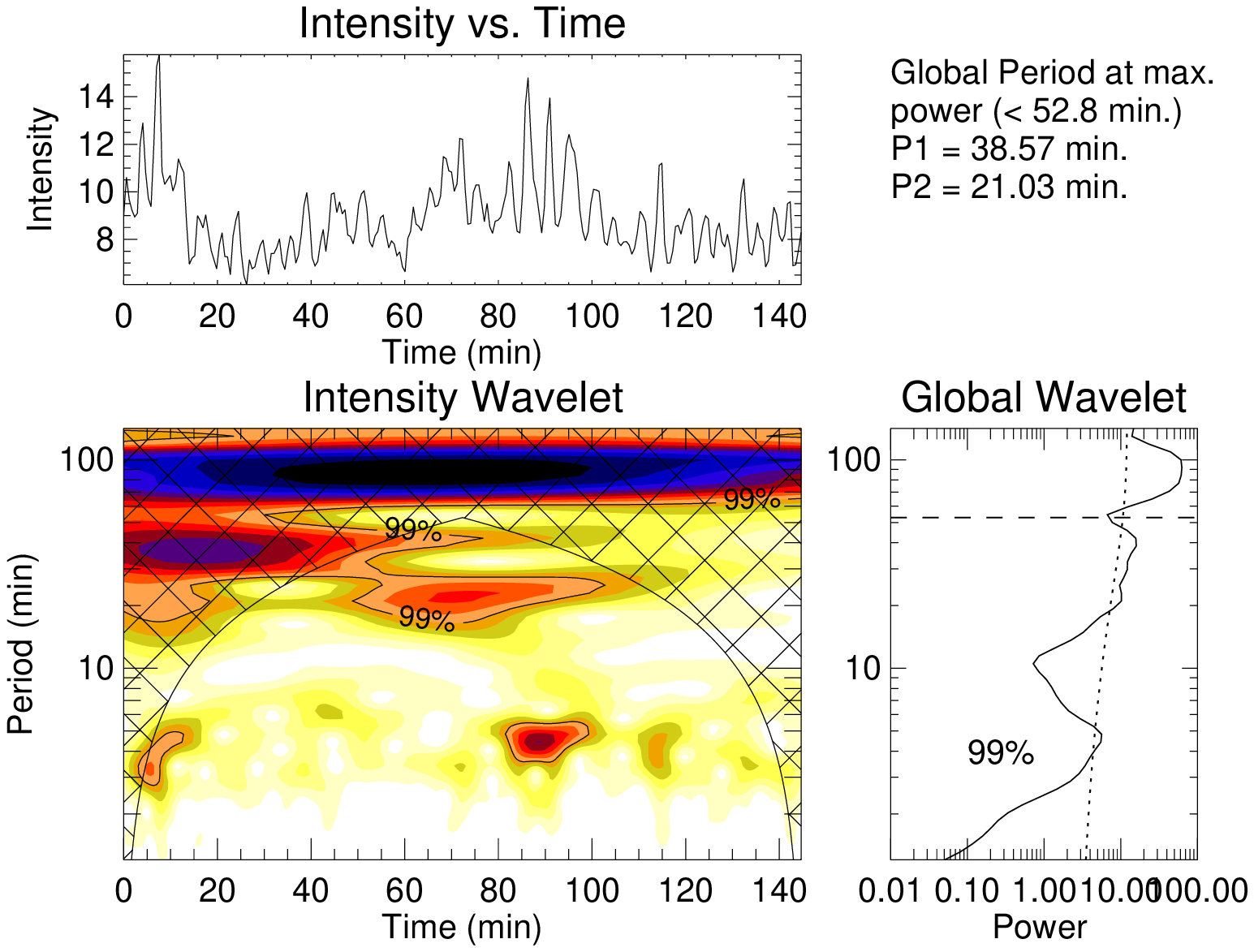}
 \caption{The wavelet results for the inter--network location as identified from TRACE 1550~\AA\ XT map
 at solar$-Y\approx 25\arcsec$ and solar$-X\approx 14\arcsec$ obtained from the
light curves of Ca~{\sc ii} H (left) and TRACE 1550~\AA\ (right). 
 In each set, the top panels show the original
light curve. The bottom left panels show the color--inverted wavelet power spectrum
 with $99~\%$ confidence level contours while
 the bottom right panels show the global (averaged over time) wavelet power spectrum with $99\%$
 global confidence level drawn. 
 The periods P1 and P2 at the locations of the first two maxima in the global wavelet spectrum are 
 printed above the global wavelet spectrum.}
 \label{fig:wav_int}
\end{figure*}

In order to focus our attention on representative network and inter--network
regions we first constructed a distance--time (XT) map for SOT images and
TRACE 1550~\AA\ images for a fixed solar$-X$ $\approx 14 \arcsec$. Corresponding Fourier power
distributions along the $Y-$direction are plotted next to the XT maps
(see Figure~\ref{fig:xt_image}). One can easily  identify the network and the
inter--network regions from the XT maps (see left panel of
Figure~\ref{fig:xt_image}). The representative network and 
inter--network regions were chosen at solar$-Y$ $\approx 18 \arcsec$ and
$\approx 25\arcsec$, respectively for a wavelet analysis.
Details on the wavelet analysis, which provides information on the temporal
variation of the signal,
are described by \inlinecite{1998BAMS...79...61T}. For the convolution with the time series in the
wavelet transform, the Morlet function was chosen.  
The wavelet analysis was applied on the light curves obtained from SOT Ca~{\sc ii}
H, and TRACE 1550~\AA\  for network and inter--network locations
and are shown in Figures~\ref{fig:wav_net} and \ref{fig:wav_int} where the top
panels show the variation of the intensity with time. The light curves were obtained from the summed 
$1\arcsec \times 1\arcsec$ areas at the locations centered around the mentioned positions.

In the wavelet spectrum, the cross-hatched regions are locations where estimates of the oscillation
period becomes unreliable and is called the cone-of-influence (COI). As a result of the COI, the
maximum measurable period is shown by a horizontal dashed line in the global wavelet plots. The
global wavelet plots are obtained by taking the mean over the wavelet time domain which is very
similar to the Fourier transform as both are giving the distribution of power with respect to
period or frequency. The periods at the locations of the first two maximum in the global wavelet
spectrum are printed above the global wavelet spectrum.

Here also the original signal was chosen without any trend subtraction to perform the wavelet 
transform. As a result of this, the peak period falls within the
COI and the confidence level of other periods outside the COI appears less. However, in all cases,
clear peaks of different periodicities are visible in all of the wavelet plots and these may become
significant upon removal of periodicities lying within the COI.   
As mentioned earlier, we have divided the period ranges as a short period 
(2--6 minutes), an intermediate period (6--15 minutes) and a longer period (above
15 minutes). From the global wavelet spectrum, the corresponding period at the first
dominant power peak is termed P1 whereas that at the second dominant power peak
is termed P2.
Based on the wavelet analysis plots and  in particular on the global wavelet
power distribution in Figures~\ref{fig:wav_net} and \ref{fig:wav_int}, we infer 
that most of the observed periods (P1 and P2) in the network region are in
the longer period range from about 20 to 40 minutes. These are the periods which
are dominating the power maps in the network regions in Figure~\ref{fig:power_map}.
Whereas, the neighbouring inter--network region scenario is slightly different.
Ca~{\sc ii} H and TRACE 1550~\AA\ show the dominant peak period of
about 4 and 20 minutes, which are similar to the power seen in the dark regions of
Figure~\ref{fig:power_map}. The longer period
 present near 45 minutes in Ca~{\sc ii} H may be due to the \textit{Hinode} orbit.

\inlinecite{1993ApJ...414..345L} found that the long--period network oscillations in the chromosphere
are not directly correlated with velocity fluctuations in the photosphere immediately underneath.
Thus, they concluded that these disturbances are either confined to the chromosphere or are 
excited by photospheric events that take place at some horizontal distance from the point of
observation. Spicules and their activities are closely related to the chromospheric network which
extend upwards across the chromosphere reaching up to the height of the lower corona having typical
life times of 5--15 minutes \cite{1972ARA&A..10...73B,2009SSRv..149..355Z}. Thus, these 
spicules could be responsible for driving these long--period network oscillations.

\begin{figure*}[htbp]
\centering
\hspace*{-1cm}
\includegraphics[width=12cm]{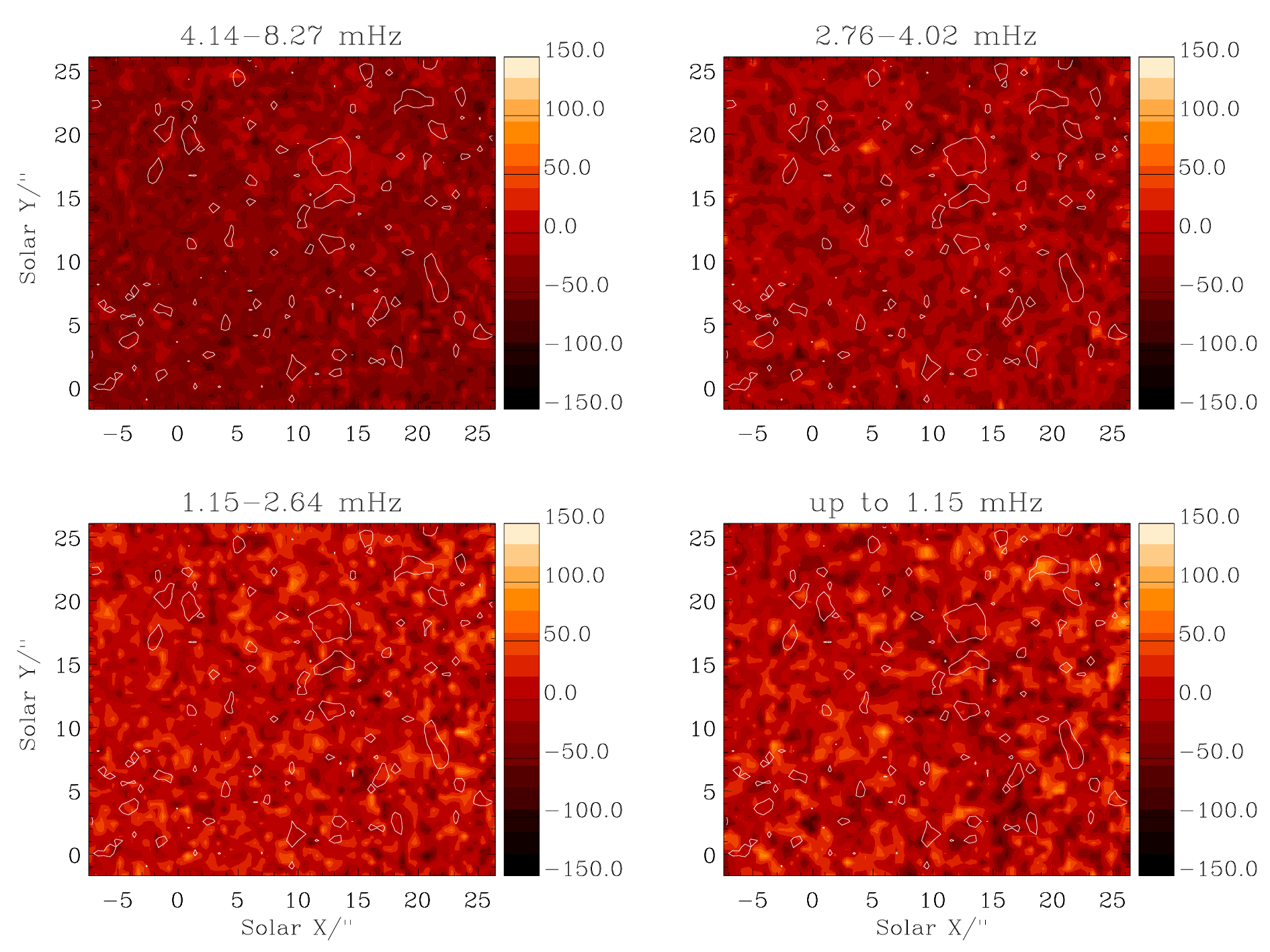}
\caption{Phase--difference maps obtained between Ca~{\sc ii}~H and TRACE 1550~\AA\ 
 on 9 April 2007 in different frequency bands as indicated on the plot. Contour
levels give the Ca~{\sc ii}~H intensity. Color scales are in degrees.} 
\label{fig:phase}
\end{figure*}

\begin{figure*}[htbp]
\centering
\hspace*{-1cm}
\includegraphics[width=12cm]{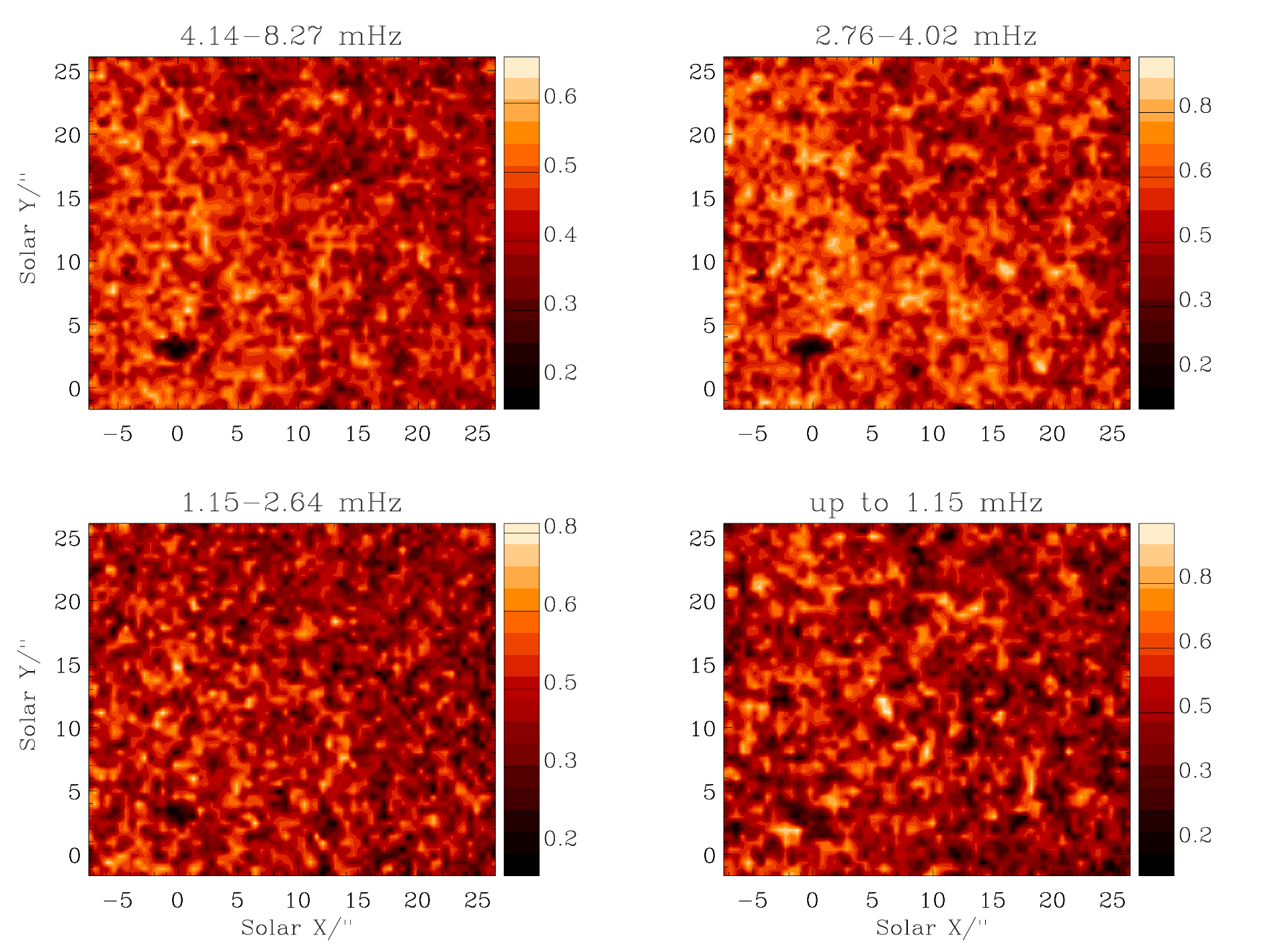}
\caption{Coherency maps obtained between Ca~{\sc ii}~H and TRACE 1550~\AA\ 
 on 9 April 2007 in different frequency bands corresponding to phase--difference maps in
Figure~\ref{fig:phase}.} 
\label{fig:coherency}
\end{figure*}

\subsection{Phase--Difference Analysis between SOT Ca~{\sc ii}~H and TRACE 1550~\AA\ Passbands}
As we see from the power maps in Figure~\ref{fig:power_map} and
oscillation periods  from the wavelet plots, the Ca~{\sc ii} H (chromosphere) and TRACE
1550~\AA\ (low transition region/chromosphere) passbands show a similar kind of oscillation.
 Hence, it is most likely that the
disturbances producing these oscillations could be due to waves propagating between the different
temperature regions as covered by
 both passbands. To investigate whether this is actually the case, we measured
the phase delays in intensity
 between the two passbands as a function of frequency for each of the measurable
pixels in the image. The SOT Ca~{\sc ii} H image array was rebinned to the spatial resolution of the
TRACE instrument and the
time series was linearly interpolated to obtain the same effective cadence as in
the TRACE 1550~\AA\ passband.

\begin{figure*}[htbp]
\centering
\hspace*{-1cm}\includegraphics[angle=90, width=13cm]{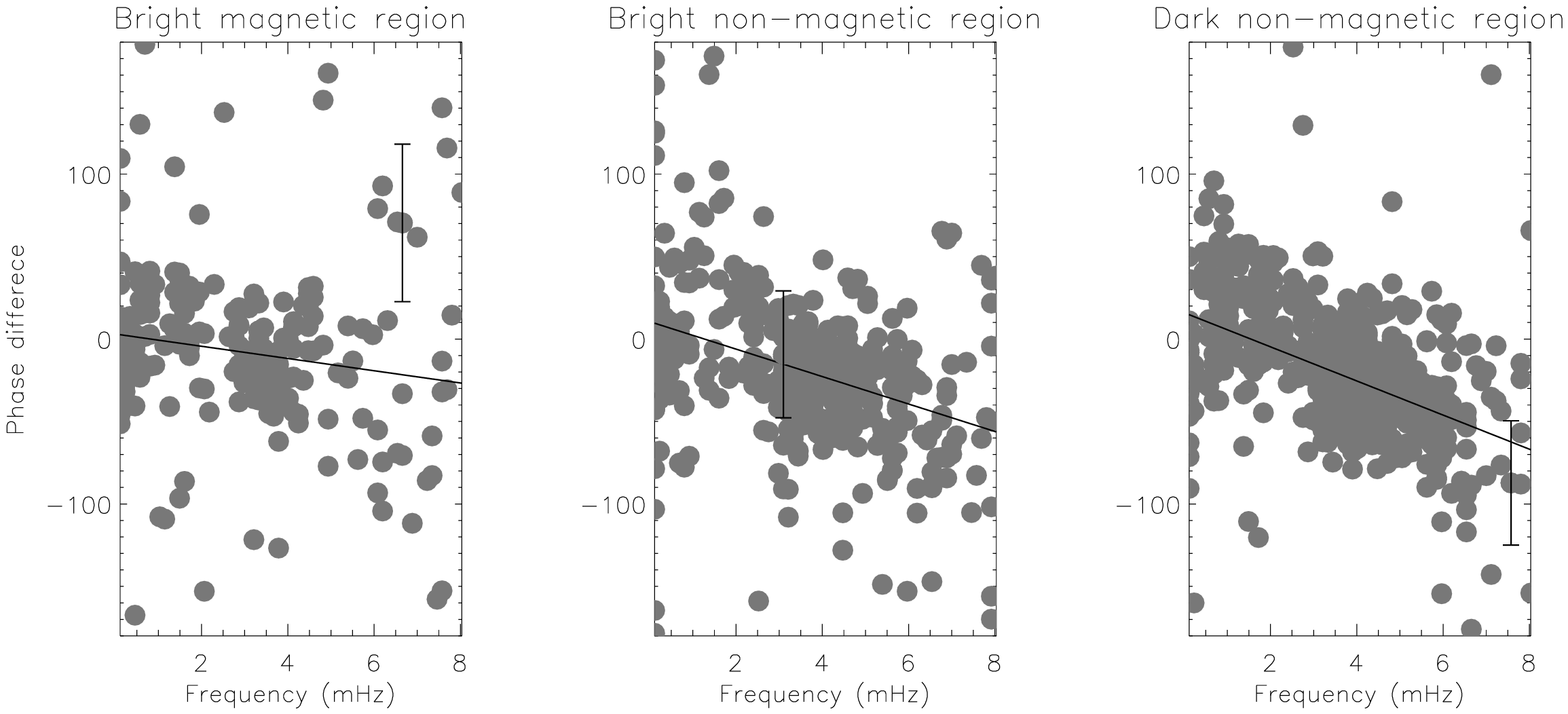}\\
\caption{Phase--distribution of Ca~{\sc ii} H and TRACE
1550~\AA\ for the selected bright magnetic, bright non-magnetic, and dark non-magnetic
regions and time delays are $-(10.3\pm3.8)$ seconds, $-(23.1\pm2.9)$ seconds,
 and $-(28.7\pm3.0)$ seconds respectively.} 
\label{fig:phase_region}
\end{figure*}
The phases were calculated from cross--power spectral estimates,
following the techniques outlined
in the appendix of \inlinecite{1999A&A...347..335D}. The errors in
the phase were calculated based on Equation (A23) of \inlinecite{1999A&A...347..335D}.
 We averaged the phases in four bands and assigned an average phase
 difference to each pixel in each band. The resultant phase--difference maps in
the four frequency  bands are shown in Figure~\ref{fig:phase} with the 
 corresponding coherency maps shown in Figure~\ref{fig:coherency}.
In the phase--difference maps, the brighter pixels (corresponding to positive phases) represent
 downward--propagating waves, whereas darker pixels (corresponding to negative phases) are representative of 
upward--propagating waves. Figure~\ref{fig:phase} clearly shows that the phase
 map for the high frequency band is globally darker
 in nature when compared to the phase maps of other lower frequency bands. The
phase map of the lowest frequency band (up to 1.15~mHz with periods
 15~minutes and above) appears to be the brightest
 as compared to the other frequency ranges globally. This indicates the
frequency dependent nature of the phase--distributions.

For high--frequency bands, the phases are more negative in nature (average phase--difference is about
 $-15\degree$) whereas for low--frequency bands, the phases are positive and near to $0\degree$\
 (average phase--difference is about $4\degree$). In Section~\ref{sec:power}, we
pointed out that the low--frequencies powers are mainly concentrated in network regions, whereas
inter--network regions have both low-- and high-- frequency powers. Hence, we interpret the negative
 phases in the high--frequency bands as due to upward--propagating
waves in the inter--network regions, whereas the near--zero positive phases in
low--frequency bands could be caused by the evanescent nature of waves present in both the network and the
inter--network regions. Also when each map is observed very closely, small mesh--like structures
are visible all over, which indicates that the upward and downward propagation occurs in the
neighbourhood, but it is difficult to identify the preferred locations of these meshes either in the
network or in the inter--network regions.

By measuring phase delays between intensity and 
velocity oscillations of different line pairs, \inlinecite{2007A&A...463..713O} 
showed  that propagating slow magnetoacoustic waves are present in coronal
holes, and that they occur preferentially in bright regions that are associated
with magnetic--field concentrations in the form of loops or bright points. In
this section, following a similar anaysis we will try to find out if there are
fixed time delays between  the  Ca~{\sc ii} H (chromosphere) and TRACE
1550~\AA\ (low transition region/chromosphere).
For a specific location, the phase--difference as a function
of frequency  is governed 
 by the equation \cite{1979ApJ...229.1147A}:
\begin{equation}
 \Delta\phi = 2\pi \textit{f}T,
\label{eq:phase}
\end{equation}
where $\textit{f}$ is the frequency and $T$ the time delay in seconds. The phase--difference will vary
 linearly with \textit{f}, and will change by $360\degree$\ over frequency intervals of 
$\Delta \textit{f} = 1/T$. This will give rise to parallel lines in $\Delta\phi$
\textit{vs.} $\textit{f}$ plots at fixed frequency intervals $(\Delta \textit{f} =1/T)$, corresponding to
a fixed time delay $T$. Hence, it is possible to measure the time delay of wave propagation between
two heights in the atmosphere for a specific location 
(\opencite{2006A&A...452.1059O}, \citeyear{2007A&A...463..713O}; \opencite{2009A&A...493..251G}).
 To obtain the time delay corresponding to different regions identified \textit{viz.}, bright magnetic,
bright non-magnetic and dark non-magnetic as marked in Figure~\ref{fig:intensity}, phase
distribution plots obtained from the Ca~{\sc ii} H and TRACE 1550~\AA\ passbands were produced for these regions.
We choose only those phases (corresponding to each spatial pixel) where the
squared coherencies are greater than a significance of 95.4\% ($2\sigma$) for further analysis.
The calculated phase delays were plotted over the full $-180\degree$ to $+180\degree$ (360$\degree$)
range and as a function of the measured oscillation frequency for all three regions 
(see Figure~\ref{fig:phase_region}).  As expected, the phases were lined up along an inclined line
as predicted by the phase equation (Equation~\ref{eq:phase}) for all three regions. A straight
line was fitted by taking into account all phase points. From the slope of this straight line,
the measured time delays are $-(10.3\pm 3.8)$ seconds, $-(23.1\pm2.9)$ seconds, and $-(28.7\pm3.0)$
 seconds for bright magnetic,
bright non-magnetic, and dark non-magnetic regions respectively. A negative time
delay indicates upward propagation of waves. In this case, the measured time delays
are very small, therefore the fixed frequency intervals $(\Delta\textit{f} =1/T)$ for the parallel 
lines are very large, and, hence, only one line is visible in the plot as
compared to the many parallel lines in previous studies 
(\opencite{2006A&A...452.1059O}, \citeyear{2007A&A...463..713O}; \opencite{2009A&A...493..251G}).
Furthermore, the slope of the fitted line is determined mostly by the high frequency phase
 difference data points, whose distance from the zero--degree line are greater than those of low--frequency
phase--difference data points. In the third panel
 of Figure~\ref{fig:phase_region} (dark non-magnetic regions), based on the scatter of the data points at
 the low--frequency range, up to 3 mHz,
 phases may be considered to be distributed around $0\degree$. This is an indication of evanescent low--frequency
 waves. At the bright magnetic regions (first panel of Figure~\ref{fig:phase_region}), there are data points
 at large phase--differences which indicate high coherency at low--frequency, and could be
 interpreted as low--frequency waves. The measured small delays between these two
passbands indicate that these two covered regions are formed very close in the 
atmosphere. Results from this analysis also indicate that the wave speed is smaller in non-magnetic regions 
as compared to magnetic regions.

\section{Summary and Conclusion}
Quiet--Sun oscillations were studied using the Fourier power and phase--difference analysis from
the intensity time series observed simultaneously in SOT Ca~{\sc ii} H and TRACE 1550~\AA.
 Fourier power maps reveal that, 3 and 5 minutes power are suppressed around the
bright magnetic--network regions at chromospheric heights. This suggests the existence of 
\textquotedblleft magnetic shadows\textquotedblright\ as seen recently by 
\inlinecite{2007A&A...461L...1V}, \inlinecite{2008A&A...488..331T}, and \inlinecite{2010A&A...510A..41K},
whereas above the 15 minutes period range we see very high--power above the magnetic network, which forms power
 halo like structure. Wavelet analysis reveals the presence of 15--30 minutes periodicities at chromospheric and 
transition--region heights (corresponding to network regions). We can only speculate on the nature of
 these low--frequency oscillations. They appear to be of slow magnetoacoustic in nature. They 
can be either propagating type (more pronounced in the network locations) or standing type 
(corresponding to the inter--network regions with low-lying magnetic loops). 

 This scenario is consistent with 
  the phase analysis results as well, where we see  the presence of pronounced upward propagation corresponding
 to magnetic--network regions and a mixture of upward and downward propagations at other locations. 
 \inlinecite{2010ApJ...713..573M}  have also reported the presence of both standing and propagating waves
 with many periods longer than 10 minutes. They also state that  a clear picture of all the wave 
modes that might be associated with active regions has not yet emerged. They further speculate that 
 the periods of the waves are related to impulsive heating which may be producing them. 
 
 Earlier work by \inlinecite{1999A&A...347..335D} indicated excess power at 
very low frequencies (based on SUMER C~{\sc ii} data), however, since these 
were sit--and--stare observations without rotational compensations, the 
implications are that this power comes from structures larger than $1\arcsec.$ 
Most earlier works were affected by the same limitation, 
either having a shorter time series or not having rotational compensation. 

Short period oscillations are dominant in the dark non-magnetic regions.
Time--delay analysis performed on phase--differences measured
between SOT Ca~{\sc ii} H and TRACE 1550 \AA\ passbands indicates different time
delays corresponding to magnetic and non-magnetic regions.   As explained by \inlinecite{2006ApJ...648L.151J},
 the low--frequency photospheric oscillations can 
propagate into the solar chromosphere through \textquotedblleft magnetoacoustic
portals\textquotedblright. They have also pointed out that a sizable fraction of
the photospheric acoustic power at frequencies below the cut-off might propagate
to higher layers within and around the quiet magnetic network elements.
In a similar way, these long period waves ($>$~5~minutes) can propagate up to coronal heights
through inclined magnetic--field lines, which could reduce the cut-off frequency \cite{2004Natur.430..536D}.
\inlinecite{2007A&A...461L...1V} showed that waves with frequencies less than the 
acoustic cut-off frequency around magnetic--network structures can 
propagate  through inclined magnetic fields in the form of fibril--like 
structures. Spicules and their activities are closely related to the 
chromospheric network and fibrils which extend upwards across the 
chromosphere reaching up to the height of the lower corona. Thus, these 
spicules could be responsible for driving the long--period network 
oscillations. Using SOT data,  \inlinecite{2010SoPh..261...35L}  showed similar
space--time distributions of intensity fluctuations from 2--3 hours sequences. In the frequency range $5.5 < f
<$ 8.0 mHz, both the G-band and Ca~{\sc ii} H-line oscillations show suppression
in the presence of magnetic fields, They also found that oscillatory powers
at these frequencies  and at lower frequencies,
lie in a mesh pattern with a cell scale of 2--3 Mm, clearly larger than normal
granulation, and with correlation times on the order of hours. In our analysis
of combined \textit{Hinode} and TRACE capabilities we find a very similar pattern. 

In open--field regions, longer periods of 10--15~minutes 
have been detected  along polar plumes \cite{1998ApJ...501L.217D,2000SoPh..196...63B,2009A&A...499L..29B}, and were
interpreted as compressive waves by \inlinecite{1999ApJ...514..441O}, 15--20 minutes period waves along
interplume regions observed by \inlinecite{2010ApJ...718...11G} were explained as Alfv\'enic or fast mode waves. 
\inlinecite{2001A&A...371.1137B} observed 2--4~mHz network oscillations in the low chromospheric and 
transition--region lines in both intensity and velocity, which were interpreted in terms of kink
and sausage waves propagating upwards along thin magnetic flux tubes.
Here, intensity oscillations may result from the presence of magnetoacoustic waves 
which could provide significant energy to heat the solar atmosphere. 

 There are a number of studies involving G--band (photospheric) and Ca~{\sc ii} H (chromospheric)
 image sequences from the \textit{Dutch Open Telescope} (DOT) \cite{2004A&A...416..333R,2005A&A...441.1183D}.
Similar studies were made for different TRACE passbands also \cite{2005A&A...430.1119D} but to 
our knowledge this is the first simultaneous  study between SOT Ca~{\sc ii} H and TRACE 1550~\AA\ passband. 

To extend this work we need spectral imaging data to identify the wave mode. This requires acquisition of long--duration 
spectral and imaging data which may be obtained from future observations such as EIS and SOT onboard
\textit{Hinode}, and AIA and HMI onboard SDO. In order to establish whether these waves are transverse or 
longitudinal to the magnetic field, observations of magnetic oscillations at varying limb distances are
needed to measure the significant horizontal components of these motions. 

\acknowledgements                 
We thank the referee for their careful reading and valuable suggestions which has enabled 
us to improve the manuscript substantially.  Research at Armagh Observatory
 is grant-aided by the N.~Ireland Dept. of Culture, Arts and Leisure. We thank STFC for 
support via ST/J001082/1. The \textit{Transition Region and Coronal Explorer} 
(TRACE), is a mission of the Stanford--Lockheed Institute for Space Research, and part of the 
NASA Small Explorer program. \textit{Hinode} is a Japanese mission developed and launched by ISAS/JAXA, 
with NAOJ as domestic partner and NASA and STFC (UK) as international partners.  
It is operated by these agencies in co-operation with ESA and NSC (Norway).

\bibliographystyle{spr-mp-sola.bst}
\bibliography{/home/girjesh/research/papers/references}

\end{article} 
\end{document}